\documentclass[twocolumn,10pt,cleanfoot]{asme2ej}

\usepackage{graphics}
\usepackage{graphicx} 
\usepackage{amsmath,bm}
\usepackage{bbm}
\usepackage{amsfonts}
\usepackage{mathtools}
\usepackage{url}
\usepackage{xcolor}
\usepackage{textcomp}
\usepackage{subcaption}
\usepackage{float}

\usepackage{booktabs}
\usepackage{siunitx}

\title{Constrained multibody kinematics and dynamics in absolute coordinates: a discussion of three approaches to representing rigid body rotation}

\author{Alexandra Kissel
}

\author{Jay Taves
}

\author{Dan Negrut\thanks{Corresponding author.}
	\affiliation{
		Department of Mechanical Engineering\\
		University of Wisconsin-Madison\\
		Madison, WI 53706\\
		\{akissel,jtaves,negrut\}@wisc.edu
	}
}

\newcommand{\reps}{\bf{r}\ensuremath{\bm{\epsilon}}}
\newcommand{\rA}{\bf{rA}}
\newcommand{\rp}{\bf{rp}}

\newcommand{\norm}[1]{{ \left| { \left| #1 \right| }  \right| }}
\newcommand{\vect}[1]{{\bf{#1}}}
\newcommand{\matr}[1]{{\bf{#1}}}
\newcommand{\constr}{{\Phi}}
\newcommand{\CFOV}{coefficients of the first order variation}

\newcommand{\oriMat}{\matr{A}}
\newcommand{\oriMatOne}[1]{\matr{A}_{#1}}
\newcommand{\oriMatTOne}[1]{\matr{A}_{#1}^T}

\newcommand{\sbelChi}{{\bar {\bm{\chi}}}}

\newcommand{\omBar}{{{\bar{\bm{\omega}}}}}
\newcommand{\omBarTilde}{{\tilde{\bar{\bm{\omega}}}}}
\newcommand{\omBarTildeB}[1]{\omBarTilde{}_{#1}}
\newcommand{\tildeEulRotVecCorr}{{\tilde{\bar{\bm{\theta}}}}}
\newcommand{\eulRotVecCorr}{{{\bar{\bm{\theta}}}}}

\newcommand{\sbar}{\bar{\vect{s}}}
\newcommand{\tildeabarBody}[1]{\tilde{\bar{\vect{a}}}_{#1}}
\newcommand{\aBody}[1]{{{\vect{a}}}_{#1}}
\newcommand{\aBarBody}[1]{{\bar{\vect{a}}}_{#1}}
\newcommand{\aBarTBody}[1]{{\bar{\vect{a}}}_{#1}^T}

\newcommand{\distBodyBody}[2]{\vect{d}_{#1 #2}}
\newcommand{\distBodyBodyT}[2]{\distBodyBody{#1}{#2}^T}

\renewcommand{\d}[1]{\ensuremath{\operatorname{d}\!{#1}}}

\newcommand{\SOrtho}{\ensuremath{\textrm{SO(3)}}}
\begin{document}
	
	\maketitle

	\begin{abstract}
	{\it
	We compare three approaches to posing the index 3 set of differential algebraic equations (DAEs) associated with the constrained multibody dynamics problem formulated in absolute coordinates. The first approach works directly with the orientation matrix and therefore eschews the need for generalized coordinates used to produce the orientation matrix $ \oriMat $. The approach is informed by the fact that rotation matrices belong to the \SOrtho{} Lie matrix group. The second approach employs Euler parameters, while the third uses Euler angles. In all cases, the index 3 DAE problem is solved via a first order implicit numerical integrator. We note a roughly twofold speedup of {\rA} over {\reps}, and a 1.2 -- 1.3 times speedup of {\reps} over {\rp}. The tests were carried out in conjunction with four 3D mechanisms. The improvements in simulation speed of the {\rA} approach are traced back to a simpler form of the equations of motion and more concise Jacobians that enter the numerical solution. The contributions made herein are twofold. First, we provide first order variations of all the quantities that enter the {\rA} formulation when used in the context of implicit integration; i.e., sensitivity of the kinematic constraints for all lower pair joints, as well as the sensitivity of the constraint reaction forces. Second, to the best of our knowledge, there is no other contribution that compares head to head the solution efficiency of {\rA}, {\rp}, and {\reps} in the context of the multibody dynamics problem posed in absolute coordinates.
	}
	\end{abstract}

	\section{Introduction}
	\label{sec:intro}
	In rigid multibody dynamics simulation using absolute coordinates, several approaches can be employed to track the attitude of a body in 3D motion. This paper is  concerned with comparing the performance (accuracy and efficiency) of three approaches. The first, called herein {\rA}, solves the kinematics and dynamics problems by computing the orientation matrix $ {\oriMat}_i \in {\mathbb{R}}^{3\times 3}$ of each body $ i $ in the mechanical system directly. The other two approaches use Euler parameters ({\rp}) and Euler angles ({\reps}) to express the orientation matrix \( {\oriMat}_i \). To the best of our knowledge, the results of such a study has not been reported in the literature.
	
	While {\rA}, {\rp}, and {\reps} use different paths to produce the orientation matrix $ \oriMat $, all three formulations share the same high-level process for generating a numerical solution. Specifically, for dynamics analysis, the numerical solution is produced using a direct index 3 approach that uses a first order implicit integrator to discretize the Newton-Euler constrained equations of motion \cite{orlandea77,BrCP96}. Herein, we compare the simulation times and the number of Newton-Raphson iterations taken until convergence of {\rA}, {\rp}, and {\reps} in both kinematics and dynamics. We also confirm the first order accuracy of the three dynamics solvers.
	
	\section{Kinematics Analysis}
	\label{sec:kinematics}	
	In multibody kinematics and dynamics analyses, the kinematic constraints modeling the lower-pair joints, e.g., revolute joint, spherical joint, hook joint, translational joint, etc., can be formulated in terms of a small set of geometric constraints (GCONs) \cite{Haug89}. The ones chosen herein are called {\textbf{DP1}}, {\textbf{DP2}}, {\textbf{D}}, and {\textbf{CD}}, and summarized in Table \ref{tab:JointsFromGCONs}.
	
	\begin{table}
		\begin{center}
			\resizebox{\columnwidth}{!}{ 
			\begin{tabular}{@{}l l l l l@{}} 
				\toprule
				& \textbf{DP1} & \textbf{DP2} & \textbf{D} & \textbf{CD} \\ 
				\midrule
				Intermediate constraint &  & & & \\ 
				\hspace{5pt} Perpendicular 1 $(\perp1)$ & $\times\times$ & & & \\ 
				\hspace{5pt} Perpendicular 2 $(\perp2)$ & & $\times\times$ & & \\ 
				Joints & & & & \\ 
				\hspace{5pt} Spherical (SJ) & & & & $\times\times\times$ \\ 
				\hspace{5pt} Universal (UJ) & $\times$ & & & $[\times\times\times]_{SJ}$ \\ 
				\hspace{5pt} Cylindrical (CJ) & $[\times\times]_{\perp1}$ & $[\times\times]_{\perp2}$ & & \\ 
				\hspace{5pt} Revolute (RJ) & $[\times\times]_{\perp1}$ & & & $[\times\times\times]_{SJ}$ \\ 
				\hspace{5pt} Translational (TJ) & $\times[\times\times]_{\perp1}$ & $[\times\times]_{\perp2}$ & & \\ 
				\bottomrule
			\end{tabular}
			}
			\caption{Summary of modeling various joints using four basic GCONs. {\textit{Perpendicular 1}} constrains a vector $\vect{c}_j$ on body \textit{j} to be perpendicular to a plane of body \textit{i} defined by two noncolinear vectors $\vect{a}_i$ and $\vect{b}_i$ (from where the two $ \times $ signs). {\textit{Perpendicular 2}} constrains a vector $\protect\overrightarrow{P_iQ_j}$ from body \textit{i} to body \textit{j} to be perpendicular to a plane defined by two noncolinear vectors $\vect{a}_i$ and $\vect{b}_i$.}
			\label{tab:JointsFromGCONs}
		\end{center}
	\end{table} 

	The {\textbf{DP1}} (``dot product 1'') GCON constrains the relative motion between two bodies by requiring that the dot product between a vector attached to body $i$ and a second vector attached to body $j$ assumes a specified value. For instance, if the vectors remain at all times perpendicular, this value should stay zero throughout the simulation. {\textbf{DP1}} has the following attributes; i.e., it requires the following pieces of information to be properly set up: $(i)$ body $i$, the associated local-reference frame L-RF$_i$, and the vector ${\bar {\bf a}}_i$ that enters the dot product condition; $(ii)$ body $j$, the associated local-reference frame L-RF$_j$, and the vector ${\bar {\bf a}}_j$ that enters the dot product condition; and, ($ iii $) the value that the dot product should assume, which is prescribed through a function $f(t)$. Most often, $f(t)=0$, which indicates that the two vectors are orthogonal. If the function $f$ depends on time, this leads to ${\constr ^{DP1}}$ being a driving (rheonomic) constraint. The {\textbf{DP1}} geometric constraint is captured in the following kinematic constraint equation:
	\begin{equation}
		\label{eq:DP1definition}
		{{\constr} ^{DP1}}(i,{\bar {\bf a}}_i,j,{\bar {\bf a}}_j, f(t)) 
		=
		{{\bar {\bf a}}_i}^T {\bf A}^T_i  {\bf A}_j {\bar {\bf a}}_j - f(t)
		=
		0 \; .
	\end{equation}		

	The {\textbf{DP2}} (``dot product 2'') GCON constrains the relative motion between two bodies by requiring that the dot product between a vector ${\vec{\bf a}}_i$ on body $i$ and a second vector ${\overrightarrow{P_i Q_j}}$ from body $i$ to body $j$ assumes a specified value. {\textbf{DP2}} has the following attributes: $(i)$ body $i$, the associated L-RF$_i$, the algebraic vector ${\bar {\bf a}}_i$, and the location ${\bar{\bf s}}^P_i$ of the point $P_i$; $(ii)$ body $j$, the associated L-RF$_j$, and the location ${\bar {\bf s}}^Q_j$ of the point $Q_j$; and, $(iii)$ the value that the dot product should assume, which is prescribed through a function $f(t)$. Most often, $f(t)=0$, which indicates that the two vectors are orthogonal. If the function $f$ depends on time, this leads to ${\constr ^{DP2}}$ being a driving (rheonomic) constraint. The {\textbf{DP2}} geometric constraint is captured in the following kinematic constraint equation:
	\vspace{-10pt}
	\begin{equation}
		\label{eq:DP2equation}
		\begin{aligned}
			{{\constr} ^{DP2}} & (i,{\bar {\bf a}}_i,{\bar {\bf s}}^P_i,j,{\bar {\bf s}}^Q_j, f(t)) 
			= {{\bar {\bf a}}_i}^T {\bf A}^T_i {\bf d}_{ij} - f(t) \\
			& =
			{{\bar {\bf a}}_i}^T {\bf A}^T_i  ({\bf r}_j + {\bf A}_j {\bar {\bf s}}^Q_j -  {\bf r}_i - {\bf A}_i {\bar {\bf s}}^P_i)  - f(t) \\
			& = 0 \; .
		\end{aligned}
	\end{equation}	
	\vspace{-10pt}
	
	The {\textbf{D}} (``distance'') GCON constrains the relative motion between two bodies by requiring that the distance between point $P$ on body $i$ and point $Q$ on body $j$ assumes a specified value strictly greater than zero. {\textbf{D}} has the following attributes: $(i)$ body $i$, the associated L-RF$_i$, and the location ${\bar {\bf s}}^P_i$ of the point $P$; $(ii)$ body $j$, the associated L-RF$_j$, and the location ${\bar{\bf s}}^Q_j$ of the point $Q$; and, $(iii)$ the value that the distance between the two points should assume, which is prescribed through the function $f(t)$. Most often, $f(t)=C^2 > 0$, which defines a scleronomic kinematic constraint (the power $2$ emphasizes that the constant function assumes a non-negative value). If $f(t)$ depends on time, this leads to ${\constr ^{D}}$ being a driving (rheonomic) constraint. The {\textbf{D}} geometric constraint is captured in the following kinematic constraint equation:
	\begin{equation}
		\label{eq:Dequation}
		\begin{aligned}
			{{\constr}^{D}}& (i,{\bar {\bf s}}^P_i,j,{\bar {\bf s}}^Q_j, f(t)) 
			= 	{\bf d}_{ij}^T {\bf d}_{ij} - f(t) \vspace{0.3cm} \\
			& =  
			({\bf r}_j + {\bf A}_j {\bar {\bf s}}^Q_j -  {\bf r}_i - {\bf A}_i {\bar {\bf s}}^P_i)^T ({\bf r}_j + {\bf A}_j {\bar {\bf s}}^Q_j -  {\bf r}_i - {\bf A}_i {\bar {\bf s}}^P_i)  - f(t) \\
			& =
			0 \; .
		\end{aligned}
	\end{equation}	
	
	The {\textbf{CD}} (``coordinate difference'') GCON constrains the relative motion between two bodies by requiring that the difference between the $x$ (or $y$ or $z$) coordinate of point $P$ on body $i$ and the $x$ (or $y$ or $z$) coordinate of point $Q$ on body $j$ assumes a specified value. {\textbf{CD}} has the following attributes: $(i)$ the coordinate ${\bf c} \in \{{\bf i} , {\bf j} , {\bf k} \}$ of interest; $(ii)$ body $i$, the associated L-RF$_i$, and the location ${\bar {\bf s}}^P_i$ of the point $P$; $(iii)$ body $j$, the associated L-RF$_j$, and the location ${\bar{\bf s}}^Q_j$ of the point $Q$; and, $(iv)$ the value that the coordinate difference should assume, which is prescribed through the function $f(t)$. Note that if $f(t)=\mbox{const.}$, ${{\constr}^{CD}}$ defines a scleronomic kinematic constraint. Otherwise, it defines a driving (rheonomic) constraint. Also, often times the second body $j$ is the ground. In this case, by convention, $j=0$ (the global reference frame G-RF is attached to body 0). The {\textbf{CD}} geometric constraint is captured in the following kinematic constraint equation:
	\begin{equation}
		\label{eq:CDequation}
		\begin{aligned}
			{\constr ^{CD}}&( {\bf c}, i,{\bar {\bf s}}^P_i,j,{\bar {\bf s}}^Q_j, f(t) )= {\bf c}^T {\bf d}_{ij} - f(t) \\
			&  = {\bf c}^T ({\bf r}_j + {\bf A}_j {\bar {\bf s}}^Q_j -  {\bf r}_i - {\bf A}_i {\bar {\bf s}}^P_i) - f(t) \\
			& =
			0 \; .
		\end{aligned}
	\end{equation}
	
	The four basic GCONs provide the basis of the kinematics analysis for all three formulations: {\rA}, {\rp}, and {\reps}. All three formulations follow the same high-level process for generating the kinematics solution. Let the set of unknowns be \( {\vect{q}}^{\bm{\alpha}} \equiv [ {\vect{r}}_1^T,\ldots, {\vect{r}}_{nb}^T, {\bm{\alpha}}_{1}^T, \ldots {\bm{\alpha}}_{nb}^T]^T \) where \( \bm{\alpha} \) acts as a placeholder for the orientation representations \( \oriMat \), \( \vect{p} \), and \( \bm{\epsilon} \) for {\rA}, {\rp}, and {\reps}, respectively, and \( nb \) is the number of bodies in the system. With \( nc \) being the number of constraints, the kinematics analysis begins by solving the set of nonlinear equations 
	\begin{equation*}
	\vspace{-5pt}
	{\constr}^{\bm{\alpha}}(\vect{r}_1,\ldots,\vect{r}_{nb}, {\bm{\alpha}}_{1},\ldots,{\bm{\alpha}}_{nb}, t)= {\bf 0}_{nc}
	\end{equation*}
	using a Newton-Raphson method with the iteration matrix \( {\matr{G}}^{\bm{\alpha}} \in {\mathbb{R}}^{nc \times nc} \) to obtain the position-level data. One can then update the expressions in the matrix \( {\matr{G}}^{\bm{\alpha}} \) and obtain the velocity- and acceleration-level data by solving the linear equations 
	\vspace{-10pt}
	\[
	\begin{split}
		{\matr{G}}^{\bm{\alpha}}{\dot{\vect{q}}}^{\bm{\alpha}} = \nu^{\bm{\alpha}} \\
		{\matr{G}}^{\bm{\alpha}}{\ddot{\vect{q}}}^{\bm{\alpha}} = \gamma^{\bm{\alpha}}
	\end{split} \; .
	\]
	\vspace{-10pt}
	
	The following sections discuss the computation of \( {\matr{G}}^{\bm{\alpha}} \), \( \nu^{\bm{\alpha}} \), and \( \gamma^{\bm{\alpha}} \) for each formulation. Since to the best of our knowledge the first order variations of \( \constr^\oriMat \) needed in the {\rA} formulation to compute these kinematic quantities have not been presented elsewhere, we begin with a detailed discussion of the {\rA} formulation and continue with a succinct overview of the {\rp} and {\reps} formulations.
	\subsection{The {\rA} Formulation}
	In the {\rA} formulation, the multibody system kinematics and dynamics problems are solved at each time step by directly computing the Cartesian position $ {\vect{r}}_i  \in {\mathbb{R}}^{3}$ and orientation matrix $ {\oriMat}_i \in {\mathbb{R}}^{3\times 3}$ of each body $ i $ in the mechanical system. Both $ {\vect{r}}_i $ and $ {\oriMat}_i $ are relative to a global, fixed reference frame. The salient point is that the approach does not rely on Euler angles, Euler parameters, or similar generalized coordinates used to express the orientation matrix $ {\oriMat}_i $ in terms thereof. Instead, the orientation matrix is generated directly by the solution process and computed such that $ {\oriMat}_i^T {\oriMat}_i = {\matr{I}}_3$; i.e., $ {\oriMat}_i \in SO(3) $.  The fact that the orientation matrix $ \oriMat $ belongs to the special orthogonal Lie matrix group \SOrtho{} \cite{MarRat94,varadarajanLieGroups2013} has rich implications. However, in this contribution the connection to the Lie group and its algebra will be almost entirely bypassed. For how this connection is exploited in constrained multibody dynamics, the interested reader is referred to \cite{mullerLie2003,roboticsLieGroup2005,brulsCardonaArnold2012,terzeLieGroup2015,arnoldBDFLie2021}.
	
	Assume a local reference frame (L-RF, also called body reference frame) is attached to a rigid body that changes its orientation in time. As spelled out in \cite{eulerRotTheorem1776}, if at time $ t_0 $ the rotation matrix for the L-RF is $ \oriMat(t_0) $, then for any time $ t > t_0 $ this matrix, and implicitly the attitude of the rigid body, can be obtained by rotating the L-RF from its pose at $ t_0 $ by a certain angle $ \chi(t) $ about a unit axis $ {\bar {\vect{u}} }(t) \in {\mathbb{R}}^3$. Throughout, an ``over-bar'' is used to indicate that a vector is expressed in the L-RF. The subscript $ i $ was dropped for brevity.
	
	The expression of the rotation matrix associated with the $ ({\bar {\vect{u}} },\chi) $ rotation is given as \cite{Haug89}
	\[
	{\matr{R}}({\bar {\vect{u}} },\chi) 
	= 
	\cos \chi {\matr{I}}_{3} 
	+ 
	(1 - \cos \chi) {\bar {\vect{u}} }{\bar {\vect{u}} }^T  
	+ 
	\sin \chi ({\tilde{{\bar {\vect{u}} }}}) \; .
	\]
	The tilde operator acts on a vector to produce the cross-product matrix; i.e., \(\vect{a} \times \vect{b} = \tilde{\vect{a}}\vect{b}\), see \cite{Haug89}. After simple manipulations, 
	\begin{equation}
		\label{eq:RodriFormula}
		{\matr{R}}({\bar {\vect{u}} },\chi) 
		= 
		{\matr{I}}_{3} 
		+ 
		\sin \chi {\tilde{{\bar {\vect{u}} }}}
		+
		(1 - \cos \chi) {\tilde{{\bar {\vect{u}} }}} {\tilde{{\bar {\vect{u}} }}}  \; ,
	\end{equation}
	and therefore, based on Rodrigues's formula \cite{rodriguesFormula1816},
	\begin{equation}
		\label{eq:rotAsExpMatrix}
		{\matr{R}}({\bar {\vect{u}} },\chi) 
		= 
		\exp({\tilde{\sbelChi}}) \; ,
	\end{equation}
	where the Euler rotation vector $ {\sbelChi} (t)\equiv \chi {\bar {\vect{u}} } $, and for a matrix $ \matr{C} \in \mathbbm{R}^{n \times n}$, the matrix exponential is defined as 
	\begin{equation}
		\label{eq:matExpDefinition}
		\exp(\matr{C}) \equiv {\matr{I}}_n + \frac{1}{1!}{\matr{C}} + \frac{1}{2!}{\matr{C}}^2 + \frac{1}{3!}{\matr{C}}^3 + \ldots \quad .
	\end{equation}
	Although the series in Eq.~(\ref{eq:matExpDefinition}) might be divergent, it always converges for skew symmetric matrices $ \matr{C} $. Moreover, if the matrix is skew-symmetric, then $ \exp({\matr{C}}) \in SO(3)$ \cite{MarRat94}. As such, by composing the two rotations -- from the global reference to the orientation at $ t_0 $, and then from $ t_0 $ to $ t $, one has that
	\begin{equation}
		\label{eq:evolutionOfA-kinematics}
		\oriMat(t) = \oriMat(t_0) \exp({\tilde{\sbelChi}}) \; .
	\end{equation}
	The new orientation matrix $ \oriMat(t) $ is orthonormal since it is the product of two orthonormal matrices. Equation~(\ref{eq:evolutionOfA-kinematics}) provides the means to evolve, incrementally, the orientation matrix in the kinematics analysis: the new matrix at time $ t $ is obtained from the rotation matrix $ \oriMat(t_0) $ at a previous time $ t_0 $ multiplied by a second orientation matrix associated with the attitude change from $ t_0 $ to $ t $. 
	\subsubsection{Carrying out the Kinematics Analysis}
	\label{subsec:KinAnalysis}
	Assume the multibody system is subject to $ nc=6nb $ kinematic constraints: some scleronomic, some rheonomic. The collection of these kinematic constraints is denoted as
	\begin{equation}
		\label{eq:kinConstrtaintsEq}
		{\constr}^\oriMat(\vect{r}_1,\ldots,\vect{r}_{nb}, {\oriMat}_1,\ldots,{\oriMat}_{nb}, t)= {\bf 0}_{6nb} \; .
	\end{equation}
	At time $ t_n $, one in a sequence of time steps $ 0<t_1<t_2<\ldots < t_{end} $, the location and orientation of the bodies are computed by solving the nonlinear system of equations
	\begin{equation}
		\label{eq:kinConstrtaintsEqDiscrtized}
		{\constr}^\oriMat({\vect{r}}_{1,n}, \ldots, {\vect{r}}_{nb,n}, {\oriMat}_{1,n},\ldots,{\oriMat}_{nb,n}, t_n)= {\bf 0}_{6nb} \; .
	\end{equation}
	For convenience, the subscript $ i $ in $ {\vect{r}}_{i,n} $ and $ {\oriMat}_{i,n} $ is dropped. With the position $ {\vect{r}}_{n-1} $ and pose $ {\matr{A}}_{n-1} $ known  at time $t_{n-1}$, the immediate goal is to compute the new position $ {\vect{r}}_{n} $ and pose $ {\matr{A}}_{n} $ at $ t_n=t_{n-1} + h $. To that end, take the initial guess of the pose to be
	\begin{equation}
		\label{eq:initialPoseGuess}
		{\matr{A}}^{(0)}_{n} = {\matr{A}}_{n-1}  \exp(h \: {\omBarTilde}_{n-1}) \;.
	\end{equation}
	Subsequently, over a series of iterations \(k\), the orientation of the body is adjusted according to 
	\begin{equation}
		\label{eq:poseRefinement}
		{\matr{A}}^{(k+1)}_{n} = {\matr{A}}^{(k)}_{n} \: \exp({\tildeEulRotVecCorr}^{(k)})  \;, \qquad k=1,2,\ldots \; ,
	\end{equation}
	where $ {\eulRotVecCorr}^{(k)} $ is an Euler rotation vector that leads to a small rotation applied at each iteration in order to improve an imperfect pose $ {\oriMat}_n^{(k)} $. The process concludes at an iteration $ K $ for which $ \norm{{\eulRotVecCorr}^{(K)}} \leq \epsilon_R $, where the positive threshold value $ \epsilon_R $ is chosen to be small enough.
	
	Note that the body's orientation matrix $ \matr{A} $ does not appear in any kinematic quantity of interest by itself. Rather, it always multiplies a vector $ {\sbar} $ expressed in the local reference frame associated with the body, as in $  \matr{A}\sbar \in {\mathbbm{R}}^3 $. Since a Newton step is employed to find $ \matr{A} $, it is important to gauge how the quantity $ \vect{s} \equiv \matr{A}\sbar$ changes when $ \matr{A} $ changes slightly; i.e., from $ \matr{A} $ to $ \matr{A} \exp({\tildeEulRotVecCorr})$, where ${\eulRotVecCorr}$ is an Euler rotation vector associated with a small rotation. 
		
	Thus, using Eqs.~(\ref{eq:rotAsExpMatrix}) and~(\ref{eq:matExpDefinition}) and limiting to linear terms, it turns out that a small rotation characterized by ${\eulRotVecCorr}$ leads to a first order variation in $ {\vect{s}} $ of the form
	\begin{equation*}
		\delta_{{\eulRotVecCorr}} \vect{s} = \delta_{{\eulRotVecCorr}} (\matr{A}{\bar {\vect{s}}}) \equiv \matr{A} \exp({\tildeEulRotVecCorr}) \sbar - \matr{A} \sbar \approx - \matr{A} {\tilde {\sbar}} \: \eulRotVecCorr \equiv {\bar {\Pi}}(\matr{A}{\bar {\vect{s}}}) \: \eulRotVecCorr\; .
	\end{equation*}
	In other words, the first order variation of $ \matr{A}{\bar {\vect{s}}} $ is obtained via the operator $ {\bar {\Pi}} $ applied to this quantity to yield
	\begin{subequations}
		\begin{equation}
			\label{eq:senstivitySwrtTheta}
			\delta_{{\eulRotVecCorr}} \vect{s}
			=
			{\bar {\Pi}}(\matr{A}{\bar {\vect{s}}}) \: \eulRotVecCorr
			=
			{\bar {\Pi}}({{\vect{s}}}) \: \eulRotVecCorr
			=
			-\matr{A} {\tilde {\sbar}} \: \eulRotVecCorr \; .
		\end{equation}
		Equation~(\ref{eq:senstivitySwrtTheta}) illustrates how the representation in the global reference frame of a local vector $ {\sbar} $ changes when the matrix $ {\bf A} $ is slightly perturbed, as done via the Euler rotation vector ${\eulRotVecCorr}$. Note the similarity in the expression of the time derivative of $ \matr{A}{\bar {\vect{s}}} $ and its first order variation when expressed via the $ {\bar {\Pi}} $ operator:
		\begin{equation}
			\label{eq:sSimilarity}
			\frac{\d{}}{\d{t}}(\matr{A}{\bar {\vect{s}}}) 
			=
			{\bar {\Pi}}(\matr{A}{\bar {\vect{s}}}) \: {\bar {\bm{\omega}}}
			\quad \leftrightarrow \quad 
			\delta_{{\eulRotVecCorr}} (\matr{A}{\bar {\vect{s}}}) = {\bar {\Pi}}(\matr{A}{\bar {\vect{s}}}) \: \eulRotVecCorr \; .
		\end{equation}
		It is also relevant to understand how the representation of a vector $ {\bf s} $ that is {\textit{fixed}} in the global reference frame changes in the local reference frame associated with a rigid body when the attitude of the body is slightly perturbed, as done via the Euler rotation vector ${\eulRotVecCorr}$:	
		\begin{equation*}
			\resizebox{\columnwidth}{!} {$
			\delta_{{\eulRotVecCorr}} {\bar{\vect{s}}} = \delta_{{\eulRotVecCorr}} ({\matr{A}}^T\vect{s}) 
			\equiv 
			[{\matr{A}}\exp({\tildeEulRotVecCorr})]^T{\vect{s}}
			-{\matr{A}}^T{\vect{s}} 
			\approx
			{\widetilde{{\matr{A}}^T{\vect{s}}}} \: \eulRotVecCorr
			\equiv
			{\bar {\Pi}}({\matr{A}}^T\vect{s})\: \eulRotVecCorr
			= 
			{\tilde{\bar {\vect{s}}}} \: \eulRotVecCorr\; .
			$}
		\end{equation*}
		In other words, the variation of $\bar {\vect{s}} $ relative to ${\eulRotVecCorr}$, is obtained as
		\begin{equation}
			\label{eq:senstivityCwrtTheta}
			\delta_{{\eulRotVecCorr}} ({\matr{A}}^T\vect{s}) 
			=
			\delta_{{\eulRotVecCorr}} {\bar{\vect{s}}}
			=
			{\bar {\Pi}}({\matr{A}}^T\vect{s})\: \eulRotVecCorr
			=
			{\bar {\Pi}}(\bar{\vect{s}})\: \eulRotVecCorr
			=
			{\widetilde{{\matr{A}}^T{\vect{s}}}} \: \eulRotVecCorr \; .
		\end{equation}
		Note the similarity in the expression of the time derivative of $ \matr{A}^T{{\vect{s}}} $ and its first order variation:	
		\begin{equation}
			\label{eq:sbarSimilarity}
			\frac{\d{}}{\d{t}}\left( \matr{A}^T{{\vect{s}}} \right)
			=
			{\bar {\Pi}}(\matr{A}^T{{\vect{s}}}) \: {\bar {\bm{\omega}}}
			\; \leftrightarrow \; 
			\delta_{{\eulRotVecCorr}} (\matr{A}^T{{\vect{s}}}) 
			= {\bar {\Pi}}(\matr{A}^T{{\vect{s}}}) \: \eulRotVecCorr
			\; .
		\end{equation}
	\end{subequations}
	One can conclude that the action of the operator $ {\bar {\Pi}} $ on $ \matr{A}{\bar {\vect{s}}} $ or $ {\matr{A}}^T\vect{s} $ is computed as follows: a time derivative of the quantity is taken, and manipulations are done to express the time derivative in the form of a product between a matrix and the angular velocity expressed in the local reference frame. This matrix represents the action of the operator $ {\bar {\Pi}} $ on the quantity of interest.
	
	In the light of this discussion, assume that corrections in the kinematics analysis are carried out like
	\begin{subequations}
		\label{subeq:NewtonStepPosKinAnalysis}
		\begin{equation}
			\label{eq:correctionsKinematics}
			{{\bf r}}^{(k+1)}_{i,n} = {{\bf r}}^{(k)}_{i,n} + {\bm{\delta}}^{(k)}_{i,r}
			\quad , \qquad
			{\bar {\bm{\theta}}}^{(k+1)}_{i,n} =  {\bar {\bm{\theta}}}^{(k)}_{i,n} + {\bm{\delta}}^{(k)}_{i,A}
			\; .
		\end{equation}
		Then, this perturbation in the position and pose of the bodies will lead to a change (variation) in the value of the kinematic constraints that is approximated as
		\begin{equation}
			\label{eq:firstOrderVarKinematics}
			{\constr}^\oriMat({\vect{r}}^{(k+1)}_n, {\matr{A}}^{(k+1)}_n, t_n) 
			-
			{\constr}^\oriMat({\vect{r}}^{(k)}_n, {\matr{A}}^{(k)}_n, t_n) 
			\approx
			{\matr{G}}^{\oriMat,(k)} {\vect{\bm{\delta}}}^{(k)}  \; ,
		\end{equation}
		where, using the notation $ {\constr}^\oriMat_{,{\vect{r}}} \equiv \partial {\constr}^\oriMat/\partial {{\vect{r}}}$ for the partial derivative,
		\begin{equation}
			\label{eq:defGandDeltaKinematics}
			{\matr{G}}^{\oriMat,(k)}
			\equiv
			\begin{bmatrix}
				{\constr}^\oriMat_{,{\vect{r}}} & \; {\bar {\Pi}}({\constr^\oriMat}) 
			\end{bmatrix}\in {\mathbbm{R}^{{6nb}\times{6nb}}}
			, \; 
			{\vect{\bm{\delta}}}^{(k)}
			\equiv
			\begin{bmatrix}
				{\bm{\delta}}^{(k)}_r \\
				{\bm{\delta}}^{(k)}_{A} 
			\end{bmatrix} \in \mathbbm{R}^{6nb} \; .
		\end{equation}	
		The correction/perturbation $ {\vect{\bm{\delta}}}^{(k)} $ will be chosen to render $ {\constr^\oriMat}({\vect{r}}^{(k+1)}_n, {\matr{A}}^{(k+1)}_n, t_n) = {\bf 0}$; i.e., it will be  computed by solving the linear system
		\begin{equation}
			\label{eq:correctionDeltaKin}
			{{\bf G}}^{\oriMat,(k)}  {\bm{\delta}}^{(k)}
			=
			-{\constr}^\oriMat({\vect{r}}^{(k)}_n, {\matr{A}}^{(k)}_n, t_n)  \; .
		\end{equation}
	\end{subequations}
	The notation $ {\bar {\Pi}}(\constr^\oriMat) $ is used since the first order variation of $ {\constr}^\oriMat $ in terms of $ {\bm{\delta}}^{(k)}_r $ and $ {\bm{\delta}}^{(k)}_{A} $ cannot be formulated by resorting to partial derivatives, which would be the case should one use Euler angles or Euler parameters.
	
	Taking a time derivative of the kinematic constraints yields the velocity kinematic constraint equations:
	\begin{subequations}
		\begin{equation}
			\label{eq:rAKinTimeDer}
			\begin{aligned}
				\frac{\mbox{d} {{\constr}^\oriMat}({\bf r}, {\matr{A}}, t)}{\mbox{d} t}
				& =
				{\constr}^\oriMat_{,{\vect{r}}} {\dot {\bf r}} + {\bar {\Pi}}(\constr^\oriMat) {\bar{\bm{\omega}}}  + {\constr}^\oriMat_{t} \\
				& =
				\left[ { {\constr}^\oriMat_{,{\vect{r}}} \;\; {\bar {\Pi}}(\constr^\oriMat) } \right]  \left[{\begin{array}{c}{\dot {\bf r}} \\ {\bar {\bm{\omega}}} \end{array} }\right] + {\constr}^\oriMat_{t} 
				=
				{\bf 0}_{6nb} \; .
			\end{aligned}
		\end{equation}
		By moving all terms that do not depend on either ${\bar {\bm{\omega}}}$ or ${\dot {\bf r}}$ to the right hand side, Eq.~(\ref{eq:rAKinTimeDer}) is written in matrix notation as
		\begin{equation}
			\label{eq:kinVelAnalysis}
			{{\bf G}^\oriMat} \left[{\begin{array}{c}{\dot {\bf r}} \\ {\bar {\bm{\omega}}} \end{array} }\right] = {\bf \nu}^\oriMat_{6nb} \; .
		\end{equation}
	\end{subequations}
	Similarly, by taking a second time derivative of the kinematic constraints and moving all terms that do not depend on either $\dot {\bar {\bm{\omega}}}$ or ${\ddot {\bf r}}$ to the right hand side, one obtains the acceleration kinematic constraint equations:
	\begin{equation}
		\label{eq:kinAccAnalysis}
		{{\bf G}^\oriMat} \left[{\begin{array}{c}{\ddot {\bf r}} \\ \dot {\bar {\bm{\omega}}} \end{array} }\right] = {\bf \gamma}^\oriMat_{6nb} \;.
	\end{equation}
	It can be concluded that the kinematics analysis is then carried out as follows: at each $ t_n $, the iterative approach in Eq.~(\ref{subeq:NewtonStepPosKinAnalysis})  is used to compute $ \vect{r}_{1,n}$, $ \vect{r}_{2,n},\ldots , \vect{r}_{nb,n}$ and orientations matrices $ \oriMat_{1,n}, \oriMat_{2,n}, \ldots, \oriMat_{nb,n} $. Based on this level zero information, one evaluates the Jacobian $ \matr{G}^\oriMat $ and uses Eq.~(\ref{eq:kinVelAnalysis}) to compute the new velocities; i.e., level one information. Finally, Eq.~(\ref{eq:kinAccAnalysis}) is used to compute the new accelerations once $ {\bf \gamma}^\oriMat_{6nb} $ is evaluated based on level zero and level one information.
	\subsubsection{Computing $ \matr{G} $, $ {\bf \nu} $, and $ {\bf \gamma} $}
	\label{subsec:GCONdiscussion}
	Subsection \S\ref{subsec:KinAnalysis} discussed how to compute variations of the quantities $ \oriMat{\bar {\vect{s}}} $ and $ {\oriMat}^T{{\vect{s}}} $ in response to a small variation in the attitude of a rigid body; i.e., a small variation in $ \oriMat $ -- see Eqs.~(\ref{eq:senstivitySwrtTheta}) and (\ref{eq:senstivityCwrtTheta}).
	The question addressed in this subsection is as follows: how can one use the variations of $ \oriMat{\bar {\vect{s}}} $ and $ {\oriMat}^T{{\vect{s}}} $ to gauge the variation of the complex kinematic constraints that come up in the economy of the multibody system kinematics and dynamics analyses. 
	Based on the expressions of the DP1, DP2, D, and CD geometric kinematic constraints defined in Section \S\ref{sec:kinematics}, the sensitivities of interest are obtained as shown in Table {\ref{tab:sensitivitiesGCONs}} \cite{TR-2020-08-JayAllieDan}.
	\begin{table}[t]
		\centering
		\resizebox{\columnwidth}{!}{ 
		\begin{tabular}{ccccc} \toprule
			GCON & $ ,{\vect{r}}_i $ &  ,$\eulRotVecCorr_i$ & $ ,{\vect{r}}_j $ &  $ ,\eulRotVecCorr_j$ \\ \midrule
			$ {DP1} $  & $ {\vect{0}_{1 \times 3}} $  &  $-{\aBarTBody{j}}{\oriMatTOne{j}}\matr{A}_{i} { {\tildeabarBody{i}}} $ & $ {\vect{0}_{1 \times 3}} $  & $ -{\aBarTBody{i}}{\oriMatTOne{i}}\matr{A}_{j} { {\tildeabarBody{j}}} $ \\
			$ {DP2}$  & $ - {{\bf a}}_i^T $  & $ {\bar {\bf a}}_i^T {\tilde {\bar {\bf s}}}^P_i - {\bf d}_{ij}^T {\bf A}_i {\tilde {\bar {\bf a}}}_i $ & $ {{\bf a}}_i^T $  & $ - {{\bf a}}_i^T  {\bf A}_j {\tilde {\bar {\bf s}}}^Q_j $ \\
			$ {D}$  & $-2 {\bf d}_{ij}^T$  & $ 2 {\bf d}_{ij}^T {\bf A}_i {\tilde {\bar {\bf s}}}^P_i $ & $2 {\bf d}_{ij}^T$  & $ - 2 {\bf d}_{ij}^T {\bf A}_j {\tilde {\bar {\bf s}}}^Q_j $ \\ 
			$ {CD}$ & $-{\bf c}^T$  & $ {\bf c}^T {\bf A}_i {\tilde {\bar {\bf s}}}^P_i $ & ${\bf c}^T$  & $ - {\bf c}^T {\bf A}_j {\tilde {\bar {\bf s}}}^Q_j $ \\ \bottomrule
		\end{tabular}
		}
		\caption{Coefficients associated with the first order variation of the basic GCONs in the {\rA} formulation.}
		\label{tab:sensitivitiesGCONs}
	\end{table}
	
	For all GCONs, the contribution \( {\bf \nu}^\oriMat \) to the right hand side of the velocity equation is 
	\begin{equation}
		\vspace{-10pt}
		{\bf \nu}^\oriMat = \frac{\partial f(t)}{\partial t} \; , 
	\end{equation}
	where $ f(t) $ is the time dependent component that shows up in the definition of the GCON. Most often, $ {\bf \nu}^\oriMat=0 $, unless the user prescribes a motion via a function $ f $ explicitly depending on time.
	
	Finally, the contribution $ \gamma^\oriMat $ of each GCON to the right hand side of the acceleration equation is computed as \cite{TR-2020-08-JayAllieDan}	
	\begin{subequations}
		\label{subeq:gammaGCONs}
		\begin{align}
			\gamma^{\oriMat, DP1} &=  -\aBarBody{j} \left(\oriMatTOne{j}\oriMatOne{i} \omBarTilde{}_i \omBarTilde{}_i + \omBarTilde{}_j \omBarTilde{}_j \oriMatTOne{j}\oriMatOne{i}\right)\aBody{i} \\ \nonumber
			& + 2\omBar{}_j^T \tildeabarBody{j} \oriMatTOne{j} \oriMatOne{i} \tildeabarBody{i} \omBar{}_i + \ddot{f}(t) \\  
			\gamma^{\oriMat, DP2}  &= 2\omBar{}_i^T\tildeabarBody{i}\oriMatTOne{i}(\vect{\dot{r}}_i - \vect{\dot{r}}_j) + 2\bar{\vect{s}}_j^{QT}\omBarTildeB{j}\oriMatTOne{j}\oriMatOne{i}\omBarTildeB{i}\aBarBody{i} \\ \nonumber
			& - \bar{\vect{s}}_i^{PT}\omBarTildeB{i}\omBarTildeB{i}\aBarBody{i}  -\bar{\vect{s}}_j^{QT}\omBarTildeB{j}\omBarTildeB{j}\oriMatTOne{j}\oriMatOne{i}\aBarBody{i} \\ \nonumber
			& - \distBodyBodyT{i}{j}\oriMatOne{i}\omBarTildeB{i}\omBarTildeB{i}\aBarBody{i} + \ddot{f}(t) \\
			\gamma^{\oriMat, D} &= 2(\dot{\vect{r}}_i - \dot{\vect{r}}_j)^T (\dot{\vect{r}}_j - \dot{\vect{r}}_i) + 2\bar{\vect{s}}_j^{QT}\omBarTildeB{j}\omBarTildeB{j}\bar{\vect{s}}_j^P \\ \nonumber
			& + 2\bar{\vect{s}}_i^{PT}\omBarTildeB{i}\omBarTildeB{i}\bar{\vect{s}}_i^{P} - 4\bar{\vect{s}}_j^{QT}\omBarTildeB{j}\oriMatTOne{j}\oriMatOne{i}\omBarTildeB{i}\bar{\vect{s}}_i^{P} \\ \nonumber
			& + 4(\dot{\vect{r}}_j - \dot{\vect{r}}_i)^T\left(\oriMatOne{j}\tilde{\bar{\vect{s}}}_j^Q \omBar{}_j - \oriMatOne{i}\tilde{\bar{\vect{s}}}_i^P\omBar{}_j\right) \\ \nonumber
			& - 2\distBodyBodyT{i}{j}\left(\oriMatOne{i}\omBarTildeB{i}\tilde{\bar{\vect{s}}}_i^P\omBar{}_i - \oriMatOne{j}\omBarTildeB{i}\tilde{\bar{\vect{s}}}_j^Q\omBar{}_j\right) + \ddot{f}(t)	\\
			\gamma^{\oriMat, CD} &= \vect{c}^T \left(\oriMatOne{i}\omBarTildeB{i}\omBarTildeB{i}\bar{\vect{s}}_i^P - \oriMatOne{j}\omBarTildeB{j}\omBarTildeB{j}\bar{\vect{s}}_j^Q\right) + \ddot{f}(t)	 	\; .
		\end{align}
	\end{subequations}	

	\subsection{The {\rp} Formulation}
	The {\rp} formulation represents the orientation of a body using Euler parameters. Thus, a system with $ nb $ bodies has $ 7nb $ unknowns:  $ {\vect{r}}_i \in \mathbb{R}^3$ and $ {\vect{p}_i = [e_{i0},e_{i1},e_{i2},e_{i3}]^T} \in \mathbb{R}^4$ for each body \( i \). The \( 7 nb \) constraint equations, $ 6nb $ kinematic constraints and $ nb $ Euler normalization constraints, assume the form $ {\hat {\constr}}^\vect{p}({\vect{r}}_1, {\vect{p}}_1 , \ldots, {\vect{r}}_{nb}, {\vect{p}}_{nb},t) = {\vect{0}_{7nb}} $, or equivalently,
	\begin{equation} \label{eq:rpKinConstraintEq}
		\begin{array}{rcl}
			\mbox{kinematic} & : &{\constr^\vect{p}}({\vect{r}}_1, {\vect{p}}_1 , \ldots, {\vect{r}}_{nb}, 	{\vect{p}}_{nb},t) =  {\vect{0}}_{6nb} \\
			\mbox{normalization} & : & 1/2 \: {\vect{p}}_i^T {\vect{p}}_i - 1/2 =0, \quad 1\leq i \leq nb \; .
		\end{array}
	\end{equation}
	The nonlinear algebraic system \( \hat {\constr}^\vect{p} = {\vect{0}_{7nb}} \) is solved via a Newton-Raphson method to get $ {\vect{r}}_i $ and $ {\vect{p}}_i $. The analysis is carried out in the same manner as the {\rA} formulation, except that the first order variations of \( \hat {\constr}^\vect{p} \) in terms of $ {\bm{\delta}}^{(k)}_r $ and $ {\bm{\delta}}^{(k)}_{p} $ can both be formulated using partial derivatives so that the Jacobian \( \matr{G}^\vect{p} = [{\hat{\constr}}^\vect{p}_{,\vect{r}} \; {\hat{\constr}}^\vect{p}_{,\vect{p}}] \). Then, the velocity and acceleration equations take the form
	\begin{subequations} \label{eq:rpKinVelocityAccelerationEq}
		\begin{align}
			{\matr{G}}^\vect{p} \left[{\begin{array}{c}{\dot {\bf r}} \\ {\dot {\bf p}} \end{array} }\right] &= {\vect{\nu}}^\vect{p} \\
			{\matr{G}}^\vect{p} \left[{\begin{array}{c}{\ddot {\bf r}} \\ {\ddot {\bf p}} \end{array} }\right] &= {\vect{\gamma}}^\vect{p} \; .
		\end{align}
	\end{subequations}
	For brevity, only the contributions of \( {\vect{\nu}}^\vect{p} \) and \( {\vect{\gamma}}^\vect{p} \) for each GCON are discussed herein. For all GCONs, the contribution \( {\vect{\nu}}^\vect{p} \) to the right hand side of the velocity equation is 
	\begin{equation}
		\vspace{-10pt}
		{\vect{\nu}}^\vect{p} = \frac{\partial f(t)}{\partial t} \; , 
	\end{equation}
	where $ f(t) $ is the time dependent component that shows up in the definition of the GCON. Most often, $ {\vect{\nu}}^\vect{p}=0 $, unless the user prescribes a motion via a function $ f $ explicitly depending on time.
	
	Some notation is introduced prior to providing the expression of the $ {\vect{\gamma}}^\vect{p}  $ terms for the four GCONs. Since $ \oriMat $ depends on $ {\vect{p}} $,  
	\begin{equation}
		\label{eq:AfromEulParams}
		{\bm{A}}({\vect{p}}) = 2\left[ {\begin{array}{*{20}{c}}
				{e_0^2 + e_1^2 - \frac{1}{2}} & \;\; {{e_1}{e_2} - {e_0}{e_3}} & \;\; {{e_1}{e_3} + {e_0}{e_2}}  \vspace{0.3cm} \\
				{{e_1}{e_2} + {e_0}{e_3}} & \;\; {e_0^2 + e_2^2 - \frac{1}{2}} & \;\; {{e_2}{e_3} - {e_0}{e_1}}  \vspace{0.3cm} \\
				{{e_1}{e_3} - {e_0}{e_2}} & \;\; {{e_2}{e_3} + {e_0}{e_1}} & \;\; {e_0^2 + e_3^2 - \frac{1}{2}}  
		\end{array}} \right] \; ,
	\end{equation}
	in a step analogous to Eq.~(\ref{eq:senstivitySwrtTheta}), for a constant position vector $ {\bar {\bf s}} $ expressed in the local reference frame, a matrix $ \matr{B} $ is defined as
	\begin{subequations}
		\label{subeqs:rpNotationHelpers}
		\begin{equation}
			\label{eq:Bmatrix}
			{\bf B}({{\bf p}}, {\bar {\bf s}}) \equiv \frac{\partial[{\bf A}({{\bf p}})\cdot {\bar {\bf s}}]}{\partial {\bf p}} \; .
		\end{equation}
		In terms of time derivatives, with $ {\bf a}_i = {\bf A}_i{\bar {\bf a}}_i $ and $ {\bf d}_{ij} = {{\bf r}}_j + {{\bf A}_j}({{\bf p}}_j){\bar {\bf s}}^Q_j -  {{\bf r}}_i - {{\bf A}_i}({{\bf p}}_i){\bar {\bf s}}^P_i $, one has that:
		\begin{align}
			\label{eq:dotBmat}
			\frac{d[{\bf B}({{\bf p}}, {\bar {\bf s}}){\dot {\bf p}}]}{d\;t} & = 
			{\bf B}({\dot{\bf p}}, {\bar {\bf s}}){\dot {\bf p}} + {\bf B}({{\bf p}}, {\bar {\bf s}}){\ddot {\bf p}} \\\
			{\dot{\bf a}}_i & = {\bf B}({\bf p}_i, {\bar {\bf a}}_i) {\dot {\bf p}}_i \\ {\ddot{\bf a}}_i & = {\bf B}({\dot{\bf p}}_i, {\bar {\bf a}}_i) {\dot {\bf p}}_i + {\bf B}({{\bf p}}_i, {\bar {\bf a}}_i) {\ddot {\bf p}}_i \\
			{\dot {\bf d}}_{ij} & = {\dot{\bf r}}_j + {\bf B}({{\bf p}}_j, {\bar {\bf s}}^Q_j){\dot {\bf p}}_j -  {\dot{\bf r}}_i - {\bf B}({{\bf p}}_i, {\bar {\bf s}}^P_i){\dot {\bf p}}_i \\
			{\ddot {\bf d}}_{ij} & = {\ddot{\bf r}}_j + {\bf B}({{\bf p}}_j, {\bar {\bf s}}^Q_j){\ddot {\bf p}}_j + {\bf B}({\dot{\bf p}}_j, {\bar {\bf s}}^Q_j){\dot {\bf p}}_j \\ \nonumber
			& - {\ddot{\bf r}}_i - {\bf B}({{\bf p}}_i, {\bar {\bf s}}^P_i){\ddot {\bf p}}_i - {\bf B}({\dot{\bf p}}_i, {\bar {\bf s}}^P_i){\dot {\bf p}}_i \; .
		\end{align}
	\end{subequations}
	Taking two time derivatives of the kinematic constraints and simple algebraic manipulations drawing on definitions in Eq.~(\ref{subeqs:rpNotationHelpers}) lead to the following  ${\gamma}^\vect{p}$ expressions:
	\begin{subequations} 
		\begin{align}
			{\gamma}^{\vect{p}, DP1} & = { {\bf a}}_i^T {\bf B}({\dot {\bf p}}_j, {\bar {\bf a}}_j){\dot {\bf p}}_j
			- {{\bf a}}_j^T {\bf B}({\dot {\bf p}}_i, {\bar {\bf a}}_i){\dot {\bf p}}_i \\ \nonumber
			& - 2 {\dot {\bf a}}_i^T {\dot {\bf a}}_j + {\ddot f}(t) \\ 
			{\gamma}^{\vect{p}, DP2} & =
			- {{\bf a}}_i^T {\bf B}({\dot {\bf p}}_j, {\bar {\bf s}}^Q_j){\dot {\bf p}}_j
			+ {{\bf a}}_i^T {\bf B}({\dot {\bf p}}_i, {\bar {\bf s}}^P_i){\dot {\bf p}}_i \\ \nonumber
			& - {{\bf d}}_{ij}^T {\bf B}({\dot {\bf p}}_i, {\bar {\bf a}}_i){\dot {\bf p}}_i
			- 2 {\dot {\bf a}}_i^T {\dot {\bf d}}_{ij} + {\ddot f}(t) \\ 
			{\gamma}^{\vect{p}, D} & =
			- 2{{\bf d}}_{ij}^T {\bf B}({\dot {\bf p}}_j, {\bar {\bf s}}^Q_j){\dot {\bf p}}_j \\  \nonumber
			& + 2{{\bf d}}_{ij}^T {\bf B}({\dot {\bf p}}_i, {\bar {\bf s}}^P_i){\dot {\bf p}}_i
			- 2 {\dot {\bf d}}_{ij}^T {\dot {\bf d}}_{ij} +  {\ddot f}(t) \\ 
			{\gamma}^{\vect{p}, CD} & =
			{{\bf c}}^T {\bf B}({\dot {\bf p}}_i, {\bar {\bf s}}^P_i){\dot {\bf p}}_i \\ \nonumber
			& - {{\bf c}}^T {\bf B}({\dot {\bf p}}_j, {\bar {\bf s}}^Q_j){\dot {\bf p}}_j  +  {\ddot f}(t) \; .
		\end{align}
	\end{subequations}
	\subsection{The {\reps} Formulation}
	The {\reps} formulation represents the orientation of a body using the triple of ZXZ-intrinsic Euler rotation angles. Thus, a system with \( nb \) bodies has \( 6nb \) unknowns:  $ {\vect{r}}_i \in \mathbb{R}^3$ and \( {\bm{\epsilon}}_i = \begin{bmatrix}\phi_i, & \theta_i, & \psi_i \end{bmatrix}^T \in \mathbb{R}^3 \) for each body \( i \). The kinematic constraint equations assume the form 
	\begin{equation} \label{eq:rEpsKinConstraintEq}
		{\constr^{\bm{\epsilon}}}({\vect{r}}_1, {\bm{\epsilon}}_1 , \ldots, {\vect{r}}_{nb}, {\bm{\epsilon}}_{nb},t) =  {\vect{0}}_{6nb} \;.
	\end{equation}
	Similar to {\rp}, the nonlinear algebraic system \( \constr^{\bm{\epsilon}} = {\vect{0}}_{6nb} \) is solved via a Newton-Raphson method to get  \( \vect{r}_i \) and \( {\bm{\epsilon}}_i \). The velocity and acceleration equations take the form
	\begin{equation} \label{eq:rEpsKinVelAccEqs}
		\begin{array}{rcl}
			{\matr{G}}^{\bm{\epsilon}} \left[{\begin{array}{c}{\dot {\bf r}} \\ {\dot {\bm{\epsilon}}} \end{array} }\right] &= {\vect{\nu}}^{\epsilon} \\ [8pt]
			{\matr{G}}^{\bm{\epsilon}} \left[{\begin{array}{c}{\ddot {\bf r}} \\ {\ddot {\bm{\epsilon}}} \end{array} }\right]	&= {\vect{\gamma}}^{\bm{\epsilon}} \; ,
		\end{array}	
	\end{equation}
	where the Jacobian \( \matr{G}^{\bm{\epsilon}} = [{\constr}^{\bm{\epsilon}}_{,\vect{r}} \; {\constr}^{\bm{\epsilon}}_{,{\bm{\epsilon}}}] \). For brevity, only the contributions of \( {\vect{\nu}}^{\bm{\epsilon}} \) and \( {\vect{\gamma}}^{\bm{\epsilon}} \) for each GCON are discussed herein. For all GCONs, the contribution \( {\vect{\nu}}^{\bm{\epsilon}} \) to the right hand side of the velocity equation is 
	\begin{equation}
		\vspace{-10pt}
		{\vect{\nu}}^{\bm{\epsilon}} = \frac{\partial f(t)}{\partial t} \; , 
	\end{equation}
	where $ f(t) $ is the time dependent component that shows up in the definition of the GCON. Most often, $ {\vect{\nu}}^{\bm{\epsilon}}=0 $, unless the user prescribes a motion via a function $ f $ explicitly depending on time.
	
	Producing $ {\vect{\gamma}}^{\bm{\epsilon}} $ requires additional notation. Dropping the body index for convenience, the single-axis rotation matrices $ {\oriMat}_1 $, $ {\oriMat}_2 $, and $ {\oriMat}_3 $ are introduced to denote the \(\phi, \theta\), and \(\psi\) rotations, respectively. Then, the orientation matrix of a given body is obtained as
	\[
	\begin{array}{rcl}
		& {\bf A}({\bm{\epsilon}}) & = {\bm{A}}_1(\phi) {\bm{A}}_2(\theta) {\bm{A}}_3(\psi)  \\ [3pt]
		& = & \! \! \! \! \! \left[ {\begin{array}{*{20}c}
				{\cos \phi } & { - \sin \phi } & 0  \\
				{\sin \phi } & {\cos \phi } & 0  \\
				0 & 0 & 1  \\
		\end{array}} \right] \! \!
		\left[ {\begin{array}{*{20}c}
				1 & 0 & 0  \\
				0 & {\cos \theta } & { - \sin \theta }  \\
				0 & {\sin \theta } & {\cos \theta }  \\
		\end{array}} \right] \! \!
		\left[ {\begin{array}{*{20}c}
				{\cos \psi } & { - \sin \psi } & 0  \\
				{\sin \psi } & {\cos \psi } & 0  \\
				0 & 0 & 1  \\
		\end{array}} \right] \; .	
	\end{array}
	\] 
	Let $ {\oriMat}_{1\phi} \equiv \partial {\oriMat}_1 /\partial \phi $, and $ {\oriMat}_{1\phi\phi} \equiv \partial^2 {\oriMat}_1 /\partial {\phi}^2 $:
	\[
	\matr{A}_{1\phi} \equiv
	\begin{bmatrix}
		-\sin{\phi} & -\cos{\phi} & 0 \\
		\cos{\phi}  & -\sin{\phi} & 0 \\
		0           & 0           & 0
	\end{bmatrix} 
	\; , \quad
	\matr{A}_{1\phi\phi} \equiv
	\begin{bmatrix}
		-\cos{\phi}  &  \sin{\phi} & 0 \\
		-\sin{\phi}  & -\cos{\phi} & 0 \\
		0           & 0            & 0
	\end{bmatrix} \;,
	\]
	with similar definitions for $ \matr{A}_{2\theta} $ and $ \matr{A}_{2\theta\theta} $, and $ \matr{A}_{3\psi} $ and $ \matr{A}_{3\psi\psi} $. Then,
	\begin{subequations}
		\begin{equation}
			\dot{\matr{A}} = \dot{\phi}\matr{A}_{1\phi}\matr{A}_2\matr{A}_3 + \dot{\theta}\matr{A}_1\matr{A}_{2\theta}\matr{A}_3 + \dot{\psi}\matr{A}_1\matr{A}_2\matr{A}_{3\psi} 
		\end{equation}
		\begin{equation}
			\label{eq:definitionD}
			\begin{aligned}
				\ddot{\matr{A}} & = \left(\ddot{\phi}\matr{A}_{1\phi} + \dot{\phi}^2\matr{A}_{1\phi\phi}\right)\matr{A}_2 \matr{A}_3 + 2\dot{\phi}\dot{\theta}\matr{A}_{1\phi}\matr{A}_{2\theta}\matr{A}_3 \\ 
				& + \matr{A}_1\left(\ddot{\theta}\matr{A}_{2\theta} + \dot{\theta}^2\matr{A}_{2\theta\theta}\right)\matr{A}_3 + 2\dot{\theta}\dot{\psi}\matr{A}_1\matr{A}_{2\theta}\matr{A}_{3\psi} \\ 
				& + \matr{A}_1 \matr{A}_2 \left(\ddot{\psi}\matr{A}_{3\psi} + \dot{\psi}^2\matr{A}_{3\psi\psi}\right) + 2\dot{\phi}\dot{\psi}\matr{A}_{1\phi}\matr{A}_2\matr{A}_{3\psi} \\
				& \equiv 
				\ddot{\phi}\matr{A}_{1\phi}\matr{A}_2 \matr{A}_3
				+ \ddot{\theta} \matr{A}_1\matr{A}_{2\theta}\matr{A}_3 
				+ \ddot{\psi} \matr{A}_1 \matr{A}_2 \matr{A}_{3\psi}
				+ {\matr{D}} \; .
			\end{aligned}	
		\end{equation}
	\end{subequations}
	Taking two time derivatives of the kinematic constraints and simple algebraic manipulations drawing on the definition of $ \matr{D} $ in Eq.~(\ref{eq:definitionD}) lead to the following $ {\vect{\gamma}}^{\bm{\epsilon}} $ expressions:
	\begin{subequations}
		\begin{align}
			\label{eq:repsGamma}
			\gamma^{{\bm{\epsilon}}, DP1} & = \aBarTBody{j} \matr{A}_j^T{{\matr{D}}_{i}} \aBarBody{i} + 2\aBarTBody{i} \dot{\matr{A}}_i^T \dot{\matr{A}}_j\aBarBody{j} \\ \nonumber
			& + \aBarTBody{i} \matr{A}_i^T {\matr{D}}_{j} \aBarBody{j} + \ddot{f}(t) \\
			\gamma^{{\bm{\epsilon}}, DP2} &= 
			-\aBarTBody{i}{\matr{D}}^T_i {\vect{d}}_{ij} - 2\aBarTBody{i}{\matr{D}}^T_i {\dot{\vect{d}}}_{ij} \\ \nonumber
			& - \aBarTBody{i} \matr{A}_i^T \left( {\matr{D}}_j \sbar_j^Q - {\matr{D}}_i \sbar_i^P \right) + \ddot{f}(t) \\
			\gamma^{{\bm{\epsilon}}, D} &= -2\dot{\vect{d}}_{ij}^T \dot{\vect{d}}_{ij} - 2\vect{d}_{ij}^T\left( {\matr{D}}_j \sbar^Q_j - {\matr{D}}_i \sbar^P_i \right) + \ddot{f}(t) \\
			\gamma^{{\bm{\epsilon}}, CD} &= {\vect{c}}^T \left( {\matr{D}}_j \sbar_j^Q - {\matr{D}}_i \sbar_i^P \right) + \ddot{f}(t) \; .
		\end{align}
	\end{subequations}
	\section{Dynamics Analysis}
	\label{sec:dynamics}
		
	\subsection{The {\rA} Formulation}
	\label{subsec:rADynamics}
	In the dynamics analysis, the evolution of the orientation matrix is governed by a differential equation, which ties the rate of change of the orientation matrix to the angular velocity of the body expressed in the body reference frame $ {\bar{\bm \omega}} \in {\mathbb{R}}^3 $ as in \cite{Haug89}
	\begin{subequations}
		\label{subeq:numericalIntegrationAmatrix}
		\begin{equation}
			\label{eq:evolutionOfA-dynamics}
			{\dot \oriMat} = \oriMat {\tilde{\bar{\bm \omega}}} \; .
		\end{equation}
		To find the evolution of $ \oriMat $ from $ t_n $ to $ t_n +h $, consider  Eq.~(\ref{eq:evolutionOfA-dynamics}) in conjunction with the Lie group version of the explicit Euler integrator, $ {\oriMat}(t) = {\oriMat}_{n+1} \exp({t{\tilde{\bar{\bm \omega}}}}_{n+1}) $ \cite{iserlesLie2000}. Setting $ t=-h $, where $ h $ is the step size, leads to $ {\oriMat}_n = {\oriMat}_{n+1} \exp({-h{\tilde{\bar{\bm \omega}}}}_{n+1}) $. Since $ \exp({\matr{C}}_1 + {\matr{C}}_2) = \exp({\matr{C}}_1) \exp({\matr{C}}_2)$ as soon as $ {\matr{C}}_1 {\matr{C}}_2 - {\matr{C}}_2 {\matr{C}}_1 = {\vect{0}}$ \cite{varadarajanLieGroups2013}, a right multiplication by $ \exp({h{\tilde{\bar{\bm \omega}}}}_{n+1})  $ leads to
		\begin{equation}
			\label{eq:evolutionOfA-NumIntegrationEuler}
			{\oriMat}_{n+1} = {\oriMat}_{n} \exp({h{\tilde{\bar{\bm \omega}}}}_{n+1}) \; ,
		\end{equation}
		with the matrix exponential evaluated based on Eqs.~(\ref{eq:rotAsExpMatrix}) and (\ref{eq:matExpDefinition}). Since $ {h{\tilde{\bar{\bm \omega}}}}_{n+1} $ is skew-symmetric, its exponential is an orthonormal matrix. As such $ {\oriMat}_{n+1} $ is a proper orthonormal matrix. Note that the Lie integration formula in Eq.~(\ref{eq:evolutionOfA-NumIntegrationEuler}) is implicit.
	\end{subequations}

	As described in \cite{Haug89}, by applying D'Alembert's principle one gets the so called Newton-Euler constrained equations of motion for body $ i $ in the system. By changing the notation in \cite{Haug89} to follow the conventions adopted herein, these equations assume the form
	\begin{subequations}
		\label{subeq:NewtonEulerEOMs}
		\begin{equation}
			\label{eq:NewtonEulerEOM}
			\left\{{ 
				\begin{array}{l} m_i {\ddot{\bf r}}_i + [{\constr^\oriMat_{,\bf r_i}]}^T {\bm \lambda} = {\bf f}_i \vspace{0.2cm} \\ 
				{\bar {\bf J}}_i {\dot{\bar {\bm{\omega}}}}_i + {\bar {\Pi}}^T_i({\constr^\oriMat}) {\bm \lambda}  = {\bar{\bm \tau}}_i 
			\end{array}}
			\right. \; , \qquad i=1,\ldots,nb\; .
		\end{equation}
		Above, $ m_i $ is body $ i $'s mass; the mass moment of inertia $ {\bar {\bf J}}_i $ is constant and diagonal; the applied force depends on level zero (position/orientation) and level one (velocity) information as in $ {\bf f}_i =  {\bf f}_i({\bf r}, {\bf A}, {\dot {\bf r}}, {\dot {\bar {\bm{\omega}}}}) \in {\mathbb{R}}^3 $; and, $ {\bar {\bm \tau}}_i \equiv {\bar {\bf n}}_i({\bf r}, {\bf A}, {\dot {\bf r}}, {{\bar {\bm{\omega}}}}) - {\tilde {\bar {\bm{\omega}}}}_i {\bar {\bf J}}_i {{\bar {\bm{\omega}}}}_i \in {\mathbb{R}}^3$ depends on the applied torque $ {\bar {\bf n}}_i  $, which is considered known and provided as a function of level zero and one information. Specifically, $ {\bar {\vect{n}}}_i$ is the sum of three components: resultant of the distributed torques $ {\bar {\vect{n}}}_i^m $, which are distributed over the volume of body $ i $; $ {\bar {\vect{n}}}_i^a$, resultant torque which is obtained from applied, pointwise torques, e.g., produced by an electric motor; and, torques induced by forces $ \vect{f}_i^P $ applied to body $ i $ at a point P of location $ {\bar {\vect{s}}}_i^P $: $ {\bar {\vect{n}}}_i^f = {\tilde{\bar {\vect{s}}}}_i^P {\oriMat}_i^T {\vect{f}}_i^P $.
		
		The motion is subject to a set of algebraic constraints
		\begin{equation}
			\label{eq:kinConstrtaintsEq4Dynamics}
			{\constr^\oriMat}(\vect{r}_1,\ldots,\vect{r}_{nb}, {\oriMat}_1,\ldots,{\oriMat}_{nb}, t)= {\bf 0}_{nc} \; ,
		\end{equation}
		where $ nc\leq 6nb $.
	\end{subequations}
	For brevity, we assume that all kinematic constraints are holonomic, yet non-holonomic constraints are handled equally well in this solution approach as long as they are linear in velocity (Pfaffian). 
	
	In the light of Eq.~(\ref{eq:evolutionOfA-NumIntegrationEuler}), the first order, implicit Euler integration scheme is posed as follows ($ h $ is the integration step size):
	\begin{subequations}
		\label{subeqs:integrationFormulas}
		\begin{align}	
			{\dot {\bf r}}_{i,n+1} & = 	{\dot {\bf r}}_{i,n} + h 	{\ddot {\bf r}}_{i,n+1} \label{eq:accTranslIntegration} \\	
			{\bar {\bm{\omega}}}_{i,n+1} & = {\bar {\bm{\omega}}}_{i,n} + h	{\dot {\bar {\bm{\omega}}}}_{i,n+1} \label{eq:accAngIntegration} \\	
			{{\bf r}}_{i,n+1} & = {{\bf r}}_{i,n} + h 	{\dot {\bf r}}_{i,n+1} \label{eq:velTranIntegration} \\	
			{{\bf A}}_{i,n+1} & = {{\bf A}}_{i,n}\exp(h {\tilde{\bar {\bm{\omega}}}}_{i,n+1}) \label{eq:AmatAngIntegration}
		\end{align}
	\end{subequations}
	The formula in Eq.~(\ref{subeqs:integrationFormulas}) is hybrid: Eqs.~(\ref{eq:accTranslIntegration}) through (\ref{eq:velTranIntegration}) are classical Backward Euler formulas; Eq.~(\ref{eq:AmatAngIntegration}) is an implicit, first order Lie integration formula. A discussion of higher order integration methods suitable for handling differential equations on $ SO(3) $ falls outside the scope of this work. The reader is also referred to \cite{brulsCardonaArnold2012,arnoldBDFLie2021} for higher order numerical integration approaches when the discussion takes place in the context of Lie groups.
	
	Given the state of the system at time $ t_n $, the immediate goal is to find the values $ {\ddot {\bf r}}_{i,n+1} $, $ {\dot{\bar {\bm{\omega}}}}_{i,n+1} $, and $ {\bm \lambda}_{n+1} $ at $ t_{n+1} $. To that end, for any set of translation and angular accelerations $ {\ddot {\bf r}}_{i,n+1} $ and $ {\dot{\bar {\bm{\omega}}}}_{i,n+1} $, one can use Eq.~(\ref{eq:accTranslIntegration}) and then Eq.~(\ref{eq:velTranIntegration}) to get the new velocity and location of the bodies, respectively; and Eq.~(\ref{eq:accAngIntegration}) and then Eq.~(\ref{eq:AmatAngIntegration}) to get the new angular velocity and orientation of the bodies. Note that the matrix exponential is computed using Rodrigues's formula in  Eq.~(\ref{eq:RodriFormula}).
	
	The values $ {\ddot {\bf r}}_{i,n+1} $, $ {\dot{\bar {\bm{\omega}}}}_{i,n+1} $, and $ {\bm \lambda}_{n+1} $ at $ t_{n+1} $ are obtained by solving (via a Newton algorithm) the discretized form of the constrained equations of motion	
	\begin{subequations}
		\label{subeq:discretizedConstrainedEOM}
		\begin{equation}
			\label{eq:discConstrainedEOMs}
			\begin{aligned} 
				{\bf M} {\ddot{\bf r}}_{n+1} + \left[[\constr^\oriMat_{,{\bf r}}]^T {\bm \lambda}\right]_{n+1} & =  {\bf f}_{n+1} \vspace{0.2cm}	\\ 
				{\bar {\bf J}} {\dot{\bar {\bm{\omega}}}}_{n+1} + \left[{\bar {\Pi}}^T({\constr^\oriMat}) {\bm \lambda}\right]_{n+1} &  =  {\bar{\bm \tau}}_{n+1} \quad , \\
				\frac{1}{h^2}{\constr^\oriMat}({\bf r}_{n+1}, {\bf A}_{n+1} , t_{n+1}) & =  {\bf 0} 
			\end{aligned}
		\end{equation}
		where the scaling by $ 1/h^2 $ is done to improve the condition number of the Newton-step Jacobian \cite{negrutHHT07} and
		\begin{equation}
			\begin{aligned} 
				{\bf M} & \equiv \mbox{diag}\{m_1 {\bf I}_3,\ldots,m_{nb} {\bf I}_3\} \\
				{\bar {\bf J}} & \equiv \mbox{diag}\{{\bar {\bf J}}_{1},\ldots,{\bar {\bf J}}_{nb}\} \\
				{\bf f} & \equiv [{\bf f}_1^T,\ldots,{\bf f}_{nb}^T]^T \\
				{\bar {\bm \tau}} & \equiv [{\bar {\bm \tau}}_1^T,\ldots,{\bar {\bm \tau}}_{nb}^T]^T \\
				{\constr^\oriMat}({\bf r}, {\bf A} , t)& \equiv {\constr^\oriMat}(\vect{r}_1,\ldots,\vect{r}_{nb}, {\oriMat}_1,\ldots,{\oriMat}_{nb}, t) \; .
			\end{aligned}
		\end{equation}
	\end{subequations}
	For 3D dynamics, $ \bf M $ and $ \bar {\bf J} $ are diagonal and constant throughout the simulation. Also, although in Eq.~(\ref{eq:discConstrainedEOMs}) the arguments of $ {\constr^\oriMat} $ are the position and orientation at time $ t_{n+1} $, note that these two quantities are functions of $ {\ddot {\bf r}}_{n+1} $ and $ {\dot{\bar {\bm{\omega}}}}_{n+1} $, respectively, through the integration formulas in Eq.~(\ref{subeqs:integrationFormulas}). Then, Eq.~(\ref{eq:discConstrainedEOMs}) can be written in concise form as 
	\begin{subequations}
		\label{subeqs:discretizedConstrEOM}
		\begin{equation}
			\label{eq:discretizedShortForm}
			{\bf g^\oriMat}({\ddot{\bf r}}_{n+1}, {\dot{\bar{\bm{\omega}}}}_{n+1}, {\bm \lambda}_{n+1})
			=
			{\bf 0}_{6nb + nc} \; ,
		\end{equation}
		where
		\begin{equation}
			\label{eq:discretizedLongForm}
			{\bf g^\oriMat}({\ddot{\bf r}}, {\dot{\bar{\bm{\omega}}}}, {\bm \lambda})
			\equiv
			\left[ {
				\begin{array}{l}
					{\bf M} {\ddot{\bf r}} - {\bf f}^c - {\bf f} \\
					{\bar {\bf J}} {\dot{\bar {\bm{\omega}}}} + {\tilde {\bar {\bm{\omega}}}} {\bar {\bf J}} {{\bar {\bm{\omega}}}} - {\bar{\bf n}}^c - {\bar {\bf n}}  \\
					\frac{1}{h^2}{\constr^\oriMat}({\bf r}, {\bf A}, t ) 		\end{array}
			}\right] \; .
		\end{equation}
		\begin{equation}
			\label{eq:notationConstrForces}
			{\bf f}^c \equiv -[\constr^\oriMat_{,{\bf r}}]^T {\bm \lambda} 
			\qquad \mbox{and} \qquad
			{\bar {\bf n}}^c \equiv -{\bar {\Pi}}^T({\constr^\oriMat}) {\bm \lambda} \; .
		\end{equation}
	\end{subequations}
	Note that $ {\bf f}^c = {\bf f}^c({\bf r}, {\bf A}, {\bm{\lambda}})$ and $ {\bar {\bf n}}^c = {\bar {\bf n}}^c({\bf r}, {\bf A}, {\bm{\lambda}})$, and both $ {\bf f}^c $ and $ {\bar {\bf n}}^c $ are linear in $ {\bm{\lambda}} $. 
	In terms of notation used, any quantity without a subscript is understood to be evaluated in the configuration associated with the time $ t_{n+1} $. If quantities that depend on the previous time steps come into play, a subscript will explicitly show when such quantities are evaluated.
	
	An iterative approach is used to solve $ {\bf g^\oriMat}({\ddot{\bf r}}, {\dot{\bar{{\bm{\omega}}}}}, {\bm \lambda}) = {\bf 0} $. The $ (k+1) $ approximation of the solution is computed by applying corrections to the iteration $ (k) $ values:
	\begin{equation}
		\label{eq:corrections}
		\begin{aligned}
			{\ddot{\bf r}}^{(k+1)} & = {\ddot{\bf r}}^{(k)} + {\bm{\delta}}^{(k)}_r \\
			{\dot{\bar {\bm{\omega}}}}^{(k+1)} & =  {\dot {\bar {\bm{\omega}}}}^{(k)} + {\bm{\delta}}^{(k)}_\omega \quad .\\
			{{\bm{\lambda}}}^{(k+1)} & = {{\bm{\lambda}}}^{(k)} +{\bm{\delta}}^{(k)}_\lambda 
		\end{aligned}	
	\end{equation}
	For notation brevity, the subscripts associated with the time step ($ n + 1$) and body index ($ i $) have been dropped above. Based on Eq.~(\ref{subeqs:integrationFormulas}), a change $ {\bm{\delta}}^{(k)}_r $ in accelerations will lead to a change $ h {\bm{\delta}}^{(k)}_r $ in velocities: $  {\dot{\bf r}}^{(k+1)} = {\dot{\bf r}}^{(k)} + h{\bm{\delta}}^{(k)}_r $; and to a change $ h^2 {\bm{\delta}}^{(k)}_r $ in positions: $  {{\bf r}}^{(k+1)} = {{\bf r}}^{(k)} + h^2 {\bm{\delta}}^{(k)}_r $. Likewise, a change $ {\bm{\delta}}^{(k)}_\omega $ in angular accelerations will lead to a change $ h {\bm{\delta}}^{(k)}_\omega $ in velocities: $  {\bar {\bm{\omega}}}^{(k+1)} = {\bar {\bm{\omega}}}^{(k)} + h {\bm{\delta}}^{(k)}_\omega $. For orientation, this type of analysis cannot be carried out relative to $ {\matr{A}} $ but rather its action on other quantities. The relevant question is as follows: in light of Eq.~(\ref{eq:AmatAngIntegration}), how does $ {\matr{A}}^{(k)} {\bar{\vect{s}}}$ change as a result of the $ {\bm{\delta}}_\omega $ change in acceleration? Thus,
	\begin{subequations}
		\label{subeq:changesDueToDDTomegaChange}
		\begin{equation}
			\label{eq:changeAbars}
			\begin{aligned}
				{\matr{A}}&^{(k+1)} {\bar{\vect{s}}} - {\matr{A}}^{(k)} {\bar{\vect{s}}} 
				\\
				= &  {\matr{A}}_n \exp( h {\tilde {\bar {\bm{\omega}}}}^{(k)} + h^2 {\tilde{\bm{\delta}}}^{(k)}_\omega) {\bar{\vect{s}}} 
				- {\matr{A}}_n \exp( h{\tilde {\bar {\bm{\omega}}}}^{(k)} ) {\bar{\vect{s}}} \\
				\approx & {\matr{A}}_n \exp( h{\tilde {\bar {\bm{\omega}}}}^{(k)})[\exp(h^2 {\tilde{\bm{\delta}}}^{(k)}_\omega) - {\matr{I}_3}] {\bar{\vect{s}}} \\
				\approx & {\matr{A}}^{(k)} h^2 {\tilde{\bm{\delta}}^{(k)}_\omega} {\bar{\vect{s}}}
				= -  h^2 {\matr{A}}^{(k)} {\tilde{{\bar{\vect{s}}}}} {\bm{\delta}}^{(k)}_\omega 
				=  h^2 \: {\bar {\Pi}}({\matr{A}}^{(k)}{\bar {\vect{s}}}) \: {\bm{\delta}}^{(k)}_\omega \; .
			\end{aligned}
		\end{equation}
		By the same token,
		\begin{equation}
			\label{eq:changeATs}
			\begin{aligned}
				[{\matr{A}}&^{(k+1)}]^T {{\vect{s}}} - [{\matr{A}}^{(k)}]^T {{\vect{s}}}  \\
				= & [{\matr{A}}_n \exp( h {\tilde {\bar {\bm{\omega}}}}^{(k)} + h^2 {\tilde{\bm{\delta}}}^{(k)}_\omega)]^T {{\vect{s}}} 
				- [{\matr{A}}_n \exp( h{\tilde {\bar {\bm{\omega}}}}^{(k)} )]^T {{\vect{s}}} \\
				= &  \exp( -h {\tilde {\bar {\bm{\omega}}}}^{(k)} - h^2 {\tilde{\bm{\delta}}}^{(k)}_\omega) {\matr{A}}_n^T {{\vect{s}}} 
				- \exp( -h{\tilde {\bar {\bm{\omega}}}}^{(k)} ){\matr{A}}_n^T {{\vect{s}}} \\
				\approx &  [\exp(-h^2 {\tilde{\bm{\delta}}}^{(k)}_\omega) - {\matr{I}_3}] \exp( -h{\tilde {\bar {\bm{\omega}}}}^{(k)}) {\matr{A}}^T_n {{\vect{s}}} \\
				\approx & - h^2 {\tilde{\bm{\delta}}^{(k)}_\omega} [{\matr{A}}^{(k)}]^T {{\vect{s}}}
				=  h^2 \: {\bar {\Pi}}([{\matr{A}}^{(k)}]^T{{\vect{s}}}) \: {\bm{\delta}}^{(k)}_\omega \; .
			\end{aligned}
		\end{equation}
	\end{subequations}
	Note that $ h^2 $ assumes small values, and also $ {{\bm{\delta}}}^{(k)}_\omega $ is typically small. As such $ h^2 {{\bm{\delta}}}^{(k)}_\omega $ is a small quantity, which justified in Eq.~(\ref{subeq:changesDueToDDTomegaChange}) making the following approximation:
	\begin{equation}
		\label{eq:approxExponential}
		\exp\left( h {\tilde {\bar {\bm{\omega}}}}^{(k)} + h^2 {\tilde{\bm{\delta}}}^{(k)}_\omega\right)
		\approx
		\exp\left( h {\tilde {\bar {\bm{\omega}}}}^{(k)}\right) \cdot \exp\left(h^2 {\tilde{\bm{\delta}}}^{(k)}_\omega\right) \; .
	\end{equation}
	The salient point is that this approximation is used to yield an iteration matrix in the Newton method. As such, it does not corrupt the underlying physics, i.e., the solution is not changed. The question answered next is this: if at iteration $ (k) $ the unknowns are updated as in Eq.~(\ref{eq:corrections}), how will the value of $ {\bf g} $ in Eq.~(\ref{eq:discretizedShortForm}) change? This variation is evaluated as
	\begin{subequations}
		\begin{equation}
			\label{eq:changeIng}
			{\bf g^\oriMat}({\ddot{\bf r}}^{(k+1)}, {\dot{\bar{\bm{\omega}}}}^{(k+1)}, {\bm \lambda}^{(k+1)}) 
			-
			{\bf g^\oriMat}({\ddot{\bf r}}^{(k)}, {\dot{\bar{\bm{\omega}}}}^{(k)}, {\bm \lambda}^{(k)}) 
			=
			{\bf G}^{\oriMat,(k)} {\bm{\delta}}^{(k)} \;,
		\end{equation}
		with $ {{\bf G}}^{\oriMat,(k)} \in \mathbbm{R}^{(6nb+nc) \times (6nb+nc)} $ and $ {\bm{\delta}}^{(k)} \in \mathbbm{R}^{6nb+nc} $ defined as
		\begin{equation}
			\label{eq:defGandDelta}
			{{\bf G}}^{\oriMat,(k)} 
			=
			\begin{bmatrix}
				{\bf G}^{\oriMat,(k)}_{rr} & {\bf G}^{\oriMat,(k)}_{r{\omega}} & {\bf G}^{\oriMat,(k)}_{r{\lambda}} \vspace{0.2cm} \\
				{\bf G}^{\oriMat,(k)}_{{\omega}r} & {\bf G}^{\oriMat,(k)}_{{\omega}{\omega}} & {\bf G}^{\oriMat,(k)}_{{\omega}{\lambda}} \vspace{0.2cm} \\
				{\bf G}^{\oriMat,(k)}_{{\lambda}r} & {\bf G}^{\oriMat,(k)}_{{\lambda}{\omega}} & {\bf 0}_{nc \times nc} 
			\end{bmatrix} \qquad
			{\bm{\delta}}^{(k)}
			=
			\begin{bmatrix}
				{\bm{\delta}}^{(k)}_r \\
				{\bm{\delta}}^{(k)}_{\omega} \\
				{\bm{\delta}}^{(k)}_{\lambda}
			\end{bmatrix} \; , 
		\end{equation}
		and the superscript $ {(k)} $ indicating that the quantities are evaluated in the configuration available at iteration $ {(k)} $. Then,
		\begin{equation}
			\label{eq:defEntriesInG}
			\begin{aligned}
				{\bf G}^{\oriMat,(k)}_{rr} & = {\bf M} - h {\bf f}_{,{\dot {\bf r}}} - h^2 ({\bf f}_{,{\bf r}} + {\bf f}^c_{,{\bf r}} )\\
				{\bf G}^{\oriMat,(k)}_{r{\omega}} & = - h {\bf f}_{,{\bar {\bm{\omega}}}} - h^2 [{\bar{\Pi}}({\bf f}) + {\bar{\Pi}}({\bf f}^c)]\\
				{\bf G}^{\oriMat,(k)}_{r{\lambda}} & = [\constr^\oriMat_{,{\bf r}}]^T\\
				{\bf G}^{\oriMat,(k)}_{{\omega}r} & = - h {\bar {\bf n}}_{,{\dot {\bf r}}} - h^2 ({\bar {\bf n}}_{,{\bf r}} + {\bar {\bf n}}^c_{,{\bf r}} )\\
				{\bf G}^{\oriMat,(k)}_{{\omega}{\omega}} & = {\bar {\bf J}} - h ({\widetilde{{\bar {\bf J}}{\bar {\bm{\omega}}}}} - {\tilde {\bar {\bm{\omega}}}}{\bar {\bf J}} + {\bar {\bf n}}_{,{\bar {\bm{\omega}}}}) - h^2 [{\bar{\Pi}}({\bar {\bf n}}) + {\bar{\Pi}}({\bar {\bf n}}^c)]\\
				{\bf G}^{\oriMat,(k)}_{{\omega}{\lambda}} & = [{\bar{\Pi}}({\constr^\oriMat})]^T\\
				{\bf G}^{\oriMat,(k)}_{{\lambda}r} & = \constr^\oriMat_{,{\bf r}}\\
				{\bf G}^{\oriMat,(k)}_{{\lambda}{\omega}} & = {\bar{\Pi}}({\constr^\oriMat}) \; .
			\end{aligned}
		\end{equation}
		Ideally, the new configuration $ (k+1) $ is a root of $ {\bf g^\oriMat} $; i.e., $ {\bf g}^{\oriMat,(k+1)} = {\bf 0} $, which leads to the Newton-step correction being computed as the solution of the linear system
		\begin{equation}
			\label{eq:compOfCorrection}
			{{\bf G}}^{\oriMat,(k)}  {\bm{\delta}}^{(k)}
			=
			-{\bf g^\oriMat}({\ddot{\bf r}}^{(k)}, {\dot{\bar{\bm{\omega}}}}^{(k)}, {\bm \lambda}^{(k)}) \; .
		\end{equation}
	\end{subequations}
	What is left at this point for the approach to be fully laid out is the computation of the sensitivities of the reaction forces and torques: $ {\bf f}^c_{,{\bf r}} $, ${\bar{\Pi}}({\bf f}^c)$, $ {\bar {\bf n}}^c_{,{\bf r}} $, and $ {\bar{\Pi}}({\bar {\bf n}}^c) $; and of the sensitivities of the applied force and torque: $ {\bf f}_{,{\bf r}} $, $ {\bar{\Pi}}({\bf f}) $, $ {\bar {\bf n}}_{,{\bf r}} $, and $ {\bar{\Pi}}({\bar {\bf n}}) $. The sensitivities of the applied forces/torques are computed on a case-by-case basis, and no general rule can be provided. However, closed form formulas can be provided for the variation of the reaction forces, see \cite{TR-2020-08-JayAllieDan}. They are reported in Table~\ref{tab:DP1reactForceSensitivities} for DP1, Table~{\ref{tab:DP2reactForceSensitivities}} for DP2, Table~\ref{tab:DreactForceSensitivities} for D, and Table~\ref{tab:CDreactForceSensitivities} for CD. Since the other lower order pairs are obtained by combining these four GCONs (see Table~\ref{tab:JointsFromGCONs}), one can assemble the {\CFOV} for the reaction force associated with high-pair joints. Each table has four rows for the {\CFOV} for the reaction forces and reaction torques associated with each joint: the reaction force $ {\bf f}_i^c $ acting on body $ i $ that enters the joint; the reaction torque $ {\bar {\bf n}}_i^c$ acting on body $ i $; the reaction force $ {\bf f}_j^c $ acting on body $ j $ that enters the joint; the reaction torque $ {\bar {\bf n}}_j^c$ acting on body $ j $. These four generalized force components are defined in Eq.~(\ref{eq:notationConstrForces}).	
	\begin{table}[t]
		\centering
		\begin{tabular}{ccccc} \toprule
			DP1 & $ ,{\vect{r}}_i $ &  ,$\eulRotVecCorr_i$ & $ ,{\vect{r}}_j $ &  $ ,\eulRotVecCorr_j$  \\ \midrule
			$ {\bf f}_i^c$  & $ {\vect{0}_{3 \times 3}} $  &  ${\vect{0}_{3 \times 3}}$ & $ {\vect{0}_{3 \times 3}} $  & $ {\vect{0}_{3 \times 3}}$ \\
			$ {\bar {\bf n}}_i^c$    & $ {\vect{0}_{3 \times 3}} $  & $\tildeabarBody{i}\widetilde{\oriMatTOne{i}\matr{A}_{j}\bar {\bf a}_j} $ & $ {\vect{0}_{3 \times 3}}$  & $-\tildeabarBody{i}\oriMatTOne{i}\matr{A}_{j}\tildeabarBody{j}$ \\
			$ {\bf f}_j^c$    & ${\vect{0}_{3 \times 3}}$  & ${\vect{0}_{3 \times 3}} $ & ${\vect{0}_{3 \times 3}}$  & ${\vect{0}_{3 \times 3}}$ \\ 
			$ {\bar {\bf n}}_j^c$ & ${\vect{0}_{3 \times 3}}$  & $ -\tildeabarBody{j}\oriMatTOne{j}\matr{A}_{i}\tildeabarBody{i}$ & ${\vect{0}_{3 \times 3}}$  & $\tildeabarBody{j}\widetilde{\oriMatTOne{j}\matr{A}_{i}\bar {\bf a}_i}$ \\ \bottomrule
		\end{tabular}
		\caption{Coefficients associated with the first order variation of the reaction force/torque for the DP1 geometric constraint. NOTE: Each table entry should be scaled by $ \lambda $, the Lagrange multiplier associated with the DP1 constraint at hand.}
		\label{tab:DP1reactForceSensitivities}
	\end{table}	
	\addtolength{\tabcolsep}{-3pt}   
	\begin{table}[t]
		\centering
		\begin{tabular}{ccccc} \toprule
			DP2 & $ ,{\vect{r}}_i $ &  ,$\eulRotVecCorr_i$ & $ ,{\vect{r}}_j $ &  $ ,\eulRotVecCorr_j$  \\ \midrule
			$ {\bf f}_i^c$  & $ {\vect{0}_{3 \times 3}} $  &  $\matr{A}_{i}\tildeabarBody{i}$ & ${\vect{0}_{3 \times 3}} $  & $ {\vect{0}_{3 \times 3}}$ \\
			$ {\bar {\bf n}}_i^c$    & $ -\tildeabarBody{i}\oriMatTOne{i} $  & $-{\tilde {\bar {\bf s}}}^P_i\widetilde{\oriMatTOne{i}(\vect{r}_j+\matr{A}_{j} { {\bar {\bf s}}}^Q_j-\vect{r}_i)} $ & $\tildeabarBody{i}\oriMatTOne{i}$  & $-\tildeabarBody{i}\oriMatTOne{i}\matr{A}_{j} {\tilde {\bar {\bf s}}}^Q_j$ \\
			$ {\bf f}_j^c$    & ${\vect{0}_{3 \times 3}}$  & $ -\matr{A}_{i}\tildeabarBody{i} $ & ${\vect{0}_{3 \times 3}}$  & ${\vect{0}_{3 \times 3}}$ \\ 
			$ {\bar {\bf n}}_j^c$ & ${\vect{0}_{3 \times 3}}$  & $ -{\tilde {\bar {\bf s}}}^Q_j\oriMatTOne{j}\matr{A}_{i}\tildeabarBody{i}$ & ${\vect{0}_{3 \times 3}}$  & ${\tilde {\bar {\bf s}}}^Q_j\widetilde{\oriMatTOne{j}\matr{A}_{i}\tildeabarBody{i}} $ \\ \bottomrule
		\end{tabular}
		\caption{Coefficients associated with the first order variation of the reaction force/torque for the DP2 geometric constraint. NOTE: Each table entry should be scaled by $ \lambda $, the Lagrange multiplier associated with the DP2 constraint at hand.}
		\label{tab:DP2reactForceSensitivities}
	\end{table}	
	\begin{table}[t]
		\centering
		\begin{tabular}{ccccc} \toprule
			D & $ ,{\vect{r}}_i $ &  ,$\eulRotVecCorr_i$ & $ ,{\vect{r}}_j $ &  $ ,\eulRotVecCorr_j$  \\ \midrule
			$ {\bf f}_i^c$  & $ {\matr{I}_{3 \times 3}} $  &  $-{\bf A}_i {\tilde {\bar {\bf s}}}^P_i$ & $ -{\matr{I}_{3 \times 3}} $  & $ {\bf A}_j {\tilde {\bar {\bf s}}}^Q_j$ \\
			$ {\bar {\bf n}}_i^c$    & $ {\tilde {\bar {\bf s}}}^P_i\oriMatTOne{i} $  & $-{\tilde {\bar {\bf s}}}^P_i\widetilde{\oriMatTOne{i}(\vect{r}_j+\matr{A}_{j} { {\bar {\bf s}}}^Q_j-\vect{r}_i)} $ & $ -{\tilde {\bar {\bf s}}}^P_i\oriMatTOne{i}$  & ${\tilde {\bar {\bf s}}}^P_i\oriMatTOne{i}{\bf A}_j {\tilde {\bar {\bf s}}}^Q_j$ \\
			$ {\bf f}_j^c$    & $-{\matr{I}_{3 \times 3}}$  & $ {\bf A}_i {\tilde {\bar {\bf s}}}^P_i $ & ${\matr{I}_{3 \times 3}}$  & $-{\bf A}_j {\tilde {\bar {\bf s}}}^Q_j$ \\ 
			$ {\bar {\bf n}}_j^c$ & $-{\tilde {\bar {\bf s}}}^Q_j\oriMatTOne{j}$  & ${\tilde {\bar {\bf s}}}^Q_j\oriMatTOne{j}{\bf A}_i {\tilde {\bar {\bf s}}}^P_i$ & ${\tilde {\bar {\bf s}}}^Q_j\oriMatTOne{j}$  & ${\tilde {\bar {\bf s}}}^Q_j\widetilde{\oriMatTOne{j}{\bf d}_{ij}} $ \\ \bottomrule
		\end{tabular}
	\caption{Coefficients associated with the first order variation of the reaction force/torque for the D geometric constraint. NOTE: Each table entry should be scaled by $ 2\lambda $, where $ \lambda $ is the Lagrange multiplier associated with the D constraint at hand.}
	\label{tab:DreactForceSensitivities}
	\end{table}
	\addtolength{\tabcolsep}{3pt}
	\begin{table}[t]
		\centering
		\begin{tabular}{ccccc} \toprule
			CD & $ ,{\vect{r}}_i $ &  ,$\eulRotVecCorr_i$ & $ ,{\vect{r}}_j $ &  $ ,\eulRotVecCorr_j$  \\ \midrule
			$ {\bf f}_i^c$  & $ {\vect{0}_{3 \times 3}} $  &  ${\vect{0}_{3 \times 3}}$ & $ {\vect{0}_{3 \times 3}} $  & $ {\vect{0}_{3 \times 3}}$ \\
			$ {\bar {\bf n}}_i^c$    & $ {\vect{0}_{3 \times 3}} $  & $-{\tilde {\bar {\bf s}}}^P_i \widetilde{\oriMatTOne{i} \bf c}$ & ${\vect{0}_{3 \times 3}}$  & ${\vect{0}_{3 \times 3}}$ \\
			$ {\bf f}_j^c$    & ${\vect{0}_{3 \times 3}}$  & ${\vect{0}_{3 \times 3}}$ & ${\vect{0}_{3 \times 3}}$  & ${\vect{0}_{3 \times 3}}$ \\ 
			$ {\bar {\bf n}}_j^c$ & ${\vect{0}_{3 \times 3}}$  & ${\vect{0}_{3 \times 3}}$ & ${\vect{0}_{3 \times 3}}$  & ${\tilde {\bar {\bf s}}}^Q_j \widetilde{\oriMatTOne{j} \bf c}$ \\ \bottomrule
		\end{tabular}
		\caption{Coefficients associated with the first order variation of the reaction force/torque for the CD geometric constraint. NOTE: Each table entry should be scaled by $ \lambda $, the Lagrange multiplier associated with the CD constraint at hand.}
		\label{tab:CDreactForceSensitivities}
	\end{table}	
	
	\subsection{The {\rp} Formulation}	
	Using the notation associated with Eq.~(\ref{eq:NewtonEulerEOM}), the equations of motion in the {\rp} formulation for body \( i \) assume the form \cite{Haug89}
	\begin{subequations}
		\begin{equation}
			\label{subeq:rpEquationsOfMotion}
			\left\{
			\begin{array}{lcl}
				m_i {\ddot{\bf r}}_i + [{\constr}^\vect{p}_{{,\bf r}_i}]^T{\bm \lambda} = {\bf f}_i \quad \vspace{0.3cm} \\
				4 [{\bf B}^\vect{p}_i]^T{\bar {\bf J}}_i  {\bf B}_i^\vect{p} {\ddot{{\vect{p}}}}_i + [{\constr}^\vect{p}_{{,\bf p}_i}]^T{\bm {\lambda}} + {\vect{p}}_i {\bm \lambda}_i^{\vect{p}} = {\hat {\bar{\bm \tau}}}_i
			\end{array}
			\right. \; , \quad i=1,\ldots,nb\; ,
		\end{equation}
		subject to a set of algebraic constraints
		\begin{equation}
			{\hat {\constr}}^\vect{p}({\vect{r}}_1, {\vect{p}}_1 , \ldots, {\vect{r}}_{nb}, {\vect{p}}_{nb},t) = {\vect{0}_{7nb}} \; ,
		\end{equation}
		where 
		\begin{equation}
			\label{eq:expressionB-pForm}
			{\matr{B}}_i^\vect{p} \equiv 
			\begin{bmatrix}
				{ - {e_{i,1}}} & {{e_{i,0}}} & {{e_{i,3}}} & { - {e_{i,2}}}  \\
				{ - {e_{i,2}}} & { - {e_{i,3}}} & {{e_{i,0}}} & {{e_{i,1}}}  \\
				{ - {e_{i,3}}} & {{e_{i,2}}} & { - {e_{i,1}}} & {{e_{i,0}}}  \\
			\end{bmatrix} \; ,
		\end{equation}
		\begin{equation}
			{\hat {\bar{\bm \tau}}}_i \equiv 2[{\bf B}^\vect{p}_i]^T ({\bar {\bf n}}^m_i + {\bar {\bf n}}_i^a) + 8 [{\dot{\bf B}}^\vect{p}_i]^T {\bar {\bf J}}_i {\dot{\bf B}}^\vect{p}_i {\vect{p}}_i \; ,
		\end{equation}
		and $ {\bm \lambda} $ and $ {\bm \lambda}_i^\vect{p} $ are the set of Lagrange multipliers associated with the geometric constraints $ {\constr}^\vect{p} $ and Euler normalization constraint $ 1/2 \: {\vect{p}}_i^T {\vect{p}}_i - 1/2 =0 $, respectively -- see Eq.~(\ref{eq:rpKinConstraintEq}).
	\end{subequations}
	
	The index 3 approach adopted here for the {\rp} formulation is discussed, for instance, in \cite{negrutJayKhude2009}. The high-level procedure is the same as discussed in section \S\ref{subsec:rADynamics} for the {\rA} formulation. Thus, only the integration scheme and discretized equations of motion unique to {\rp} are stated here. The implicit Euler integration scheme takes the form 
	\begin{equation}
		\label{eq:rpIntegrationFormulas}
		\begin{aligned}	
			{\dot{\vect{r}}}^{(k)}_{n+1} &= {\dot{\vect{r}}}_{n} + h{\ddot{\vect{r}}}^{(k)}_{n+1}  \\	
			{\dot{\vect{p}}}^{(k)}_{n+1} &= {\dot{\vect{p}}}_{n} + h{\ddot{\vect{p}}}^{(k)}_{n+1} \\	
			{{\vect{r}}}^{(k)}_{n+1} &= {{\vect{r}}}_{n} + h{\dot{\vect{r}}}^{(k)}_{n+1} \\	
			{{\vect{p}}}^{(k)}_{n+1} &= {{\vect{p}}}_{n} + h{\dot{\vect{p}}}^{(k)}_{n+1} \; .
		\end{aligned} 
	\end{equation}
	For brevity, the subscript \( n+1 \) is dropped and the following quantities are understood to be evaluated in the configuration associated with the time $ t_{n+1} $. The discretized equations of motion assume the form $ \vect{g}^\vect{p}\left( \ddot{\vect{r}}, \ddot{\vect{p}}, \bm \lambda \right) = {\vect{0}}_{8nb+nc} $, where  
	\begin{equation}
		\label{eq:discretizedEOMrp}
		{\vect{g}}^\vect{p}\left( \ddot{\vect{r}}, \ddot{\vect{p}}, \bm \lambda \right) \equiv \begin{bmatrix}
			\matr{M}\ddot{\vect{r}} + \left[{\constr}^\vect{p}_{,\vect{r}}\right]^T {\bm \lambda} - {\vect{f}}(\dot{\vect{q}}^\vect{p}, {\vect{q}}^\vect{p}, t) \\
			{\bar {\matr{J}}}^{\vect{p}}\ddot{\vect{p}} + \left[{\constr}^\vect{p}_{,\vect{p}}\right]^T {\bm \lambda} + \matr{P}^T{\bm \lambda}^{\vect{p}} - \hat{\bm \tau}({\dot{\vect{q}}}^\vect{p}, {\vect{q}}^\vect{p}, t) \\
			\frac{1}{\beta_0^2 h^2} {\hat {\constr}}^{\vect{p}}
		\end{bmatrix} \; .
	\end{equation}
	In Eq.~(\ref{eq:discretizedEOMrp}), \( {\bar {\matr{J}}}^\vect{p} \equiv \text{diag}\{\left[{\bf B}_1^{\vect{p}}\right]^T {\bar {\bf J}}_1  {\bf B}_1^{\vect{p}},\ldots, \left[{\bf B}_{nb}^{\vect{p}}\right]^T {\bar {\bf J}}_{nb}  {\bf B}_{nb}^{\vect{p}}\} \) and \( {\matr{P}} \equiv \text{diag}\{ {\vect{p}}_1^T,\ldots,{\vect{p}}_{nb}^T\} \). At time $ t_{n+1} $, the unknowns $ {\ddot{\vect{r}}} $, $ {\ddot{\vect{p}}} $, $ {\bm \lambda} $, and $ {\bm \lambda}^\vect{p} $ are solved for using an iterative Newton algorithm with the Jacobian
	\begin{equation}
	\label{subeq:JacobiansRP}
		{\bf G}^\vect{p} \equiv 
		\begin{bmatrix}
			\frac{\partial {\vect{g}}^\vect{p}}{\partial \ddot{\vect{r}}} & \frac{\partial {\vect{g}}^\vect{p}}{\partial \ddot{\vect{p}}} & \frac{\partial {\vect{g}}^\vect{p}}{\partial \lambda} & \frac{\partial {\vect{g}}^\vect{p}}{\partial {\bm \lambda}^\vect{p}}
		\end{bmatrix} \; .
	\end{equation}	
	
	\subsection{The \reps{} Formulation}
	\label{subsec:repsDynamics}
	Using again the notation associated with Eq.~(\ref{eq:NewtonEulerEOM}), the equations of motion in the {\reps} formulation assume the form \cite{TR-2020-08-JayAllieDan}
	\begin{subequations}
		\label{subeq:rEpsEquationsOfMotion}
		\begin{equation}
			\left\{
			\begin{array}{lcl}
				m_i {\ddot{\bf r}}_i + [{\constr}^{\bm{\epsilon}}_{{,\bf r}_i}]^T{\bm \lambda} & = & {\bf f}_i \quad \vspace{0.3cm} \\
				\left[{\bf B}_i^{\bm{\epsilon}}\right]^T {\bar {\bf J}}_i  {\bf B}_i^{\bm{\epsilon}} {\ddot{\bm{\epsilon}}}_i + [{\constr}^{\bm{\epsilon}}_{{,\bm{\epsilon}}_i}]^T{\bm {\lambda}} & = & {\breve {\bar{\bm \tau}}}_i \; ,
			\end{array}
		\right. \; , \quad i=1,\ldots,nb\; ,
		\end{equation}
		where 
		\begin{equation}
			\label{eq:expressionB-rEpsForm}
			{\matr{B}}_i^{\bm{\epsilon}} \equiv 
			\begin{bmatrix} 
				\sin \psi_i \sin \theta_i & \cos\psi_i & 0 \\ 
				\cos\psi_i\sin \theta_i & -\sin \psi_i & 0 \\ 
				\cos\theta_i & 0 & 1
			\end{bmatrix}  \; , 
		\end{equation}
		and
		\begin{equation}
			{\breve{\bar {\bm \tau}}}^{\bm{\epsilon}}_i = [{{\bf B}}^{\bm{\epsilon}}_i]^T
			\left(
			{\bar {\bf n}}_i
			-
			{\widetilde{{{\bf B}}^{\bm{\epsilon}}_i {\dot {\bm{\epsilon}}}_i} \: {\bar {\bf J}}_i} {{\bf B}}^{\bm{\epsilon}}_i {{\dot {\bm{\epsilon}}}_i 
			- 
			{\bar {\bf J}}_i} {\dot {{\bf B}}^{\bm{\epsilon}}_i} {{\dot {\bm{\epsilon}}}_i}
			\right) \; .
		\end{equation}
	\end{subequations}	
	The implicit Euler integration scheme takes the same form as Eq.~(\ref{eq:rpIntegrationFormulas}) with Euler angles replacing the Euler parameters. The discretized equations of motion assume the form \( {\vect{g}}^{\bm{\epsilon}}\left( {\ddot{\vect{r}}}, {\ddot{{\bm{\epsilon}}}}, \bm \lambda \right) = {\vect{0}}_{6nb+nc} \), where
	\begin{equation}
		\label{eq:discretizedEOMs}
		\begin{aligned}
			{\vect{g}}^{{\bm{\epsilon}}}\left( {\ddot{\vect{r}}}, {\ddot{{\bm{\epsilon}}}}, \bm \lambda \right) &\equiv \begin{bmatrix}
				\matr{M}\ddot{\vect{r}} + \left[{\constr}^{\bm{\epsilon}}_{\vect{r}}\right]^T \bm \lambda - {\vect{f}}({\dot{\vect{q}}}^{\bm{\epsilon}}, {\vect{q}}^{\bm{\epsilon}}, t) \\
				{\bar {\matr{J}}}^{\bm{\epsilon}}{\ddot{\bm{\epsilon}}} + \left[{\constr}^{\bm{\epsilon}}_{\bm{\epsilon}}\right]^T \bm \lambda - \breve{\bm \tau}({\dot{\vect{q}}}^{\bm{\epsilon}}, {\vect{q}}^{\bm{\epsilon}}, t) \\
				\frac{1}{\beta_0^2 h^2} {\constr}^{\bm{\epsilon}}
			\end{bmatrix} \; .
		\end{aligned}
	\end{equation}
	In Eq.~(\ref{eq:discretizedEOMs}), \( {\bar {\matr{J}}}^{\bm{\epsilon}} \equiv \text{diag}\{\left[{\bf B}_1^{\bm{\epsilon}}\right]^T {\bar {\bf J}}_1  {\bf B}_1^{\bm{\epsilon}},\ldots, \left[{\bf B}_{nb}^{\bm{\epsilon}}\right]^T {\bar {\bf J}}_{nb}  {\bf B}_{nb}^{\bm{\epsilon}}\} \). At time $ t_{n+1} $, the unknowns $ {\ddot{\vect{r}}} $, $ {\ddot{\bm{\epsilon}}} $, and $ {\bm \lambda} $ are solved for using an iterative Newton algorithm with the Jacobian
	\begin{equation}
		{\bf G}^{\bm{\epsilon}} \equiv 
		\begin{bmatrix}
			\frac{\partial {\vect{g}}^{\bm{\epsilon}}}{\partial \ddot{\vect{r}}} & \frac{\partial {\vect{g}}^{\bm{\epsilon}}}{\partial \ddot{\bm{\epsilon}}} & \frac{\partial {\vect{g}}^{\bm{\epsilon}}}{\partial {\bm \lambda}} 
		\end{bmatrix} \; .
	\end{equation}

	\section{Numerical Experiments}
	\label{sec:numExperiments}
	\subsection{Generalities}
	The performance of {\rA}, {\rp}, and {\reps} was evaluated using two simulation engines independently developed in Python by the first two co-authors. Note that production codes seeking maximum performance should use a compiled language and leverage high-performance computing techniques. As such, the particular absolute run-times achieved in this performance analysis are not remarkable in themselves and are only compared to each other. The Python codes, named C1 and C2, are available in a public git repository~\cite{metadataASME-rA-formulation2021} that includes the model definitions for all mechanisms considered herein. 
	
	Both C1 and C2 follow the kinematics and dynamics solution methods as outlined in sections~\S\ref{sec:kinematics} and~\S\ref{sec:dynamics}, respectively. Within a particular codebase, C1 or C2, the three formulations are coded using the same structure and Python libraries so that the observed speedups are not due to differences in implementation. Results from the two independently developed codes C1 and C2 confirm that the reported speedups are consistent despite their different implementations and raw runtimes.	
	
	\subsection{Mechanical systems considered}\label{subsec:mechanismsConsidered}
	The numerical experiments consider the following mechanisms modeled in 3D: single pendulum, double pendulum, slider crank, and four link. The systems have zero degrees of freedom with the exception of the double pendulum, which has two degrees of freedom. The basic GCONs used among these four mechanisms are summarized in Table~\ref{tab:mech-cons}; note that each GCON is exercised at least once. The gravitational acceleration used in all systems is \SI{-9.81}{m/s^2}. 
	\begin{table}
		\centering
		\begin{tabular}{rllll}
			\toprule
			Mechanism & CD & DP1 & DP2 & D \\ \midrule
			Single Pendulum & 3 & 3 & 0 & 0 \\
			Four Link & 12 & 6 & 0 & 0 \\
			Slider Crank & 6 & 7 & 4 & 1 \\ \bottomrule
		\end{tabular}
		\caption{Summary of the basic constrains used in each mechanism. Of particular note is that the slider crank mechanism uses all four basic constraints.}\label{tab:mech-cons}
	\end{table}
	
	The single pendulum consists of a slender rod of length \SI{4}{m} and mass \SI{78}{kg} that starts at an angle of \SI{45}{\degree} to the horizontal. The mass moment of inertia is (in SI units)
	\begin{equation*}
		{\bar {\bf J}} = \begin{bmatrix}
			0.0325 & 0 & 0 \\
			0 & 104 & 0 \\
			0 & 0 & 104
		\end{bmatrix} \; .
	\end{equation*}

	The double pendulum mechanism, the only one with excess degrees of freedom, consists of two slender rods. The first rod is \SI{4}{m} long with a mass of \SI{78}{kg}, and the second rod is \SI{2}{m} long with a mass of \SI{39}{kg}. The mass moments of inertia are (in SI units)
	\begin{equation*}
		{\bar {\bf J}}_{1} = \begin{bmatrix}
			0.0325 & 0 & 0 \\
			0 & 104 & 0 \\
			0 & 0 & 104
		\end{bmatrix}, \;{\bar {\bf J}}_{2} = \begin{bmatrix}
			0.01625 & 0 & 0 \\
			0 & 13.01 & 0 \\
			0 & 0 & 13.01
		\end{bmatrix} \; .
	\end{equation*}
	Both bodies are initially at rest. The first rod is connected to the ground with a revolute joint and positioned perpendicular to gravity along the global \(y\)-axis; the second body is connected to the first body with a revolute joint and positioned parallel to gravity along the global \(z\)-axis. In this way, all but two degrees of freedom of the bodies are constrained, with the pair swinging in a plane.
	 
\begin{figure}
	\centering
	\includegraphics*[width=\columnwidth]{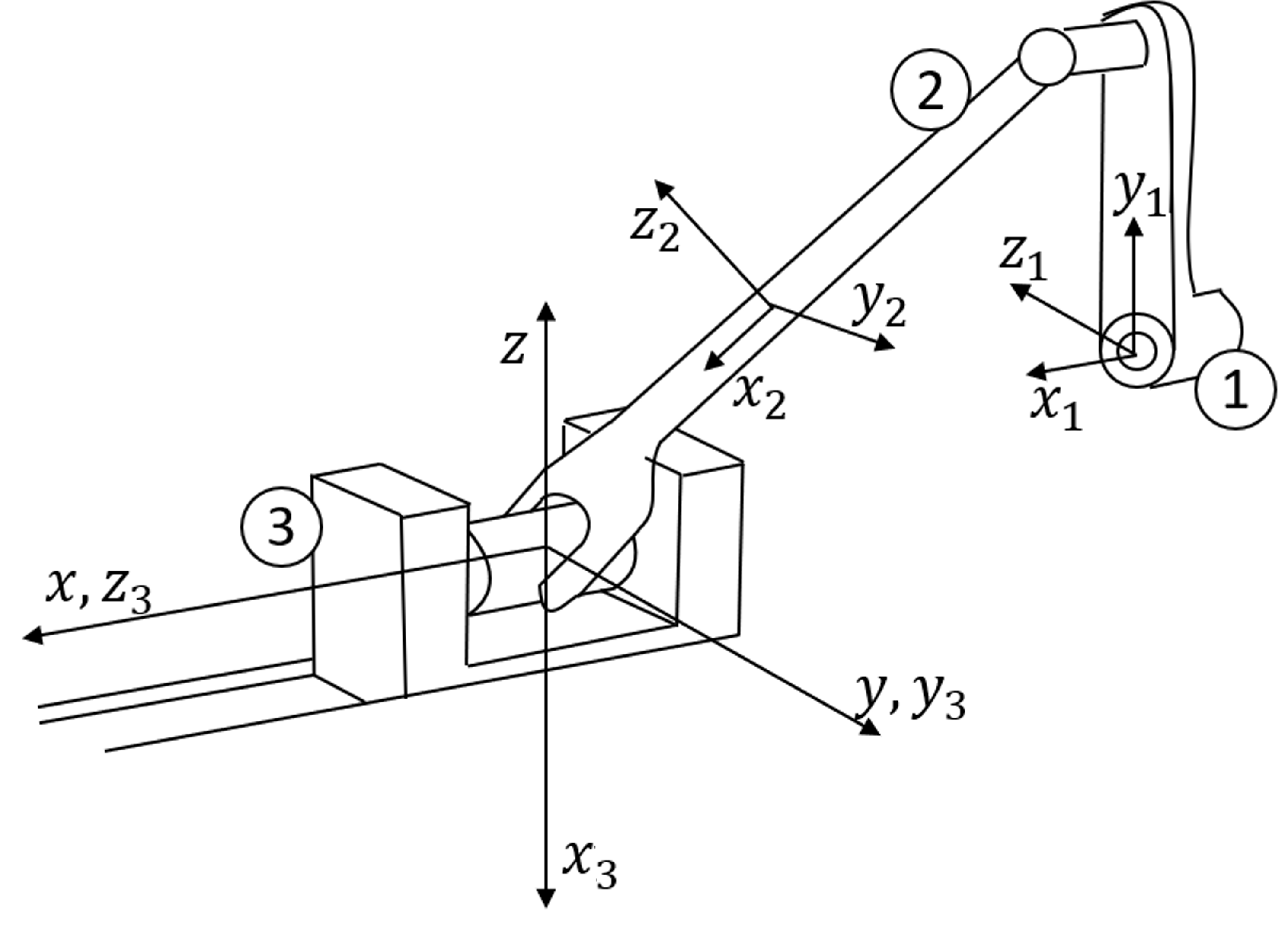}
	\caption{3D slider crank; for its parameters, along with those for the four link mechanism, see chapter 10 of \cite{Haug89}.}\label{fig:slider-crank}
	\end{figure}

	The four link and slider crank mechanisms are both closed-loop mechanisms and the model parameters are those from Chapter 10  in \cite{Haug89}. In particular, the starting positions, masses, and dimensions are the same as given therein. A schematic of the slider crank is provided in Fig.~\ref{fig:slider-crank}.
	
	A rheonomic DP1 constraint is applied to the single pendulum, slider crank, and four link mechanism to prescribe their motion. On the single pendulum, the angle between the local \(y\)-axis of the pendulum and the global negative \(z\)-axis is prescribed to be \(\frac{\pi}{2} + \frac{\pi}{4}\cos{2t}\). On the slider crank, the crank is prescribed to rotate about its negative \(x\)-axis at a rate of $ \SI[parse-numbers=false]{2\pi}{rad/s} $. For the four link mechanism, the rotor is prescribed to rotate at a rate of $ \SI[parse-numbers=false]{\pi}{rad/s} $ about its \(z\)-axis. 

	\subsection{Kinematics Analysis}
	The kinematics analyses of the zero-degree-of-freedom mechanisms are used to two ends: compare how fast the {\rA}, {\rp}, and {\reps} formulations are in carrying out the kinematics analysis; and obtain ``ground truth'' data subsequently used in an order analysis of the dynamics solver. 
	Solving the systems of linear equations in the position, velocity, and acceleration solution stages of the kinematics analysis is a critical computational bottleneck. Due to the Euler normalization constraints, the {\rp} kinematics analysis solves a linear system larger than {\rA} and {\reps} by the number of bodies in the model. As such, longer computational time for {\rp} is expected. While {\reps} has the same number of equations as {\rA}, the Jacobian of the former involves several double products of expensive trigonometric functions. 
	
	The run-times of both the {\rA} and {\reps} formulations are compared against the {\rp} formulation, the latter providing the baseline. The results shown in Table~\ref{tab:kin-speed-up} suggest that the {\rA} implementation is roughly two times faster than {\rp} and {\reps}. The performance of {\reps} is more comparable to {\rp} with roughly a 1.3x speedup, indicating that solving a larger linear system in {\rp} is counterbalanced by a more expensive way of obtaining the orientation matrix in {\reps}.
	
	\begin{table}[]
		\centering
		\begin{tabular}{rllll}
			\toprule
			& \multicolumn{2}{c}{C1} & \multicolumn{2}{c}{C2} \\
			\addlinespace[-1ex]
			& {\rA}    & {\reps} & {\rA} & {\reps} \\ \midrule
			1-Pendulum   & 2.21 & 1.40 & 2.63  & 1.27 \\
			Four Link    & 2.38 & 1.23 & 2.55  & 1.10 \\
			Slider Crank & 2.70 & 1.47 & 2.62  & 1.23 \\ \bottomrule
		\end{tabular}
		\caption{Kinematics analysis speedup, relative to {\rp} formulation; step size \num{1e-3} s. Position analysis stopped when the norm of the correction was less than \num{1e-10}. All times were computed as the average across 10 runs of the simulation.}\label{tab:kin-speed-up}
	\end{table}

\begin{figure}
		\centering
		\includegraphics*[width=\columnwidth]{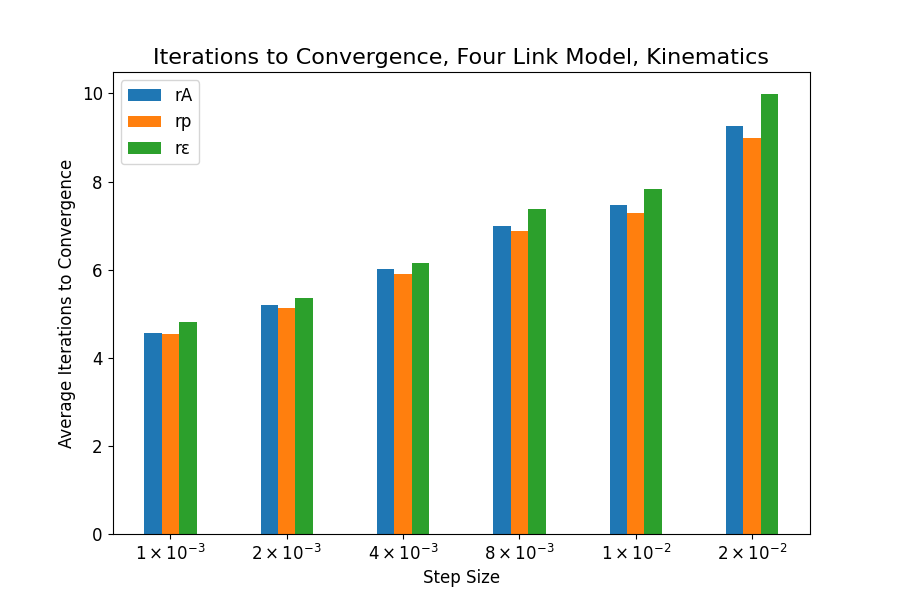}
		\caption{four link mechanism, kinematics: Iterations to convergence for position analysis. Stopping criteria of \num{1e-11}. Iterations reported are the average across a \SI{3}{s} long analysis. The dependence on step-size is not surprising; note, however, that there is no significant difference between the three formulations.}\label{fig:four link-iter-kin}
	\end{figure}

	It is also insightful to confirm that the number of iterations to convergence during the position analysis is comparable across formulations. As shown in Table~\ref{tab:kin-iters}, {\rA}, {\rp}, and {\reps} perform similarly by this metric, requiring less iterations for the simple single pendulum than the more complex four link and slider crank. Note that the number of iterations to convergence is dependent on the step-size but not significantly dependent on the formulation used, see Fig.~\ref{fig:four link-iter-kin}. 
	
	Ultimately, the results of the kinematics analysis suggest the following observations: $(i)$ {\rA} is roughly twice as fast as {\reps} and {\rp}; and $(ii)$ the number of iterations in the kinematics position analysis is comparable for {\rA}, {\rp}, and {\reps}. 
	
	\begin{table}[]
		\centering
		\begin{tabular}{rllllll}
			\toprule
			& \multicolumn{3}{c}{C1} & \multicolumn{3}{c}{C2} \\
			\addlinespace[-1ex]
			& {\rp} & {\rA}    & \reps{} & {\rp} & {\rA} & \reps{} \\ \midrule
			1-Pendulum   & 3.82 & 3.80 & 3.80 & 3.82 & 3.80 & 3.80 \\
			Four Link    & 4.47 & 4.50 & 4.77 & 4.47 & 4.50 & 4.77 \\
			Slider Crank & 4.83 & 4.70 & 4.79 & 4.83 & 4.70 & 4.79 \\ \bottomrule
		\end{tabular}
		\caption{Iterations to convergence, kinematics, position analysis. Step size of \num{1e-3}, stopping criteria of \num{1e-10}. Iterations shown here are the average across an entire run of the simulation (typically at least one full period of the driven body).}
		\label{tab:kin-iters}
	\end{table}
	
	\subsection{Dynamics Analysis}
\begin{figure}
		\centering
		\includegraphics*[width=\columnwidth]{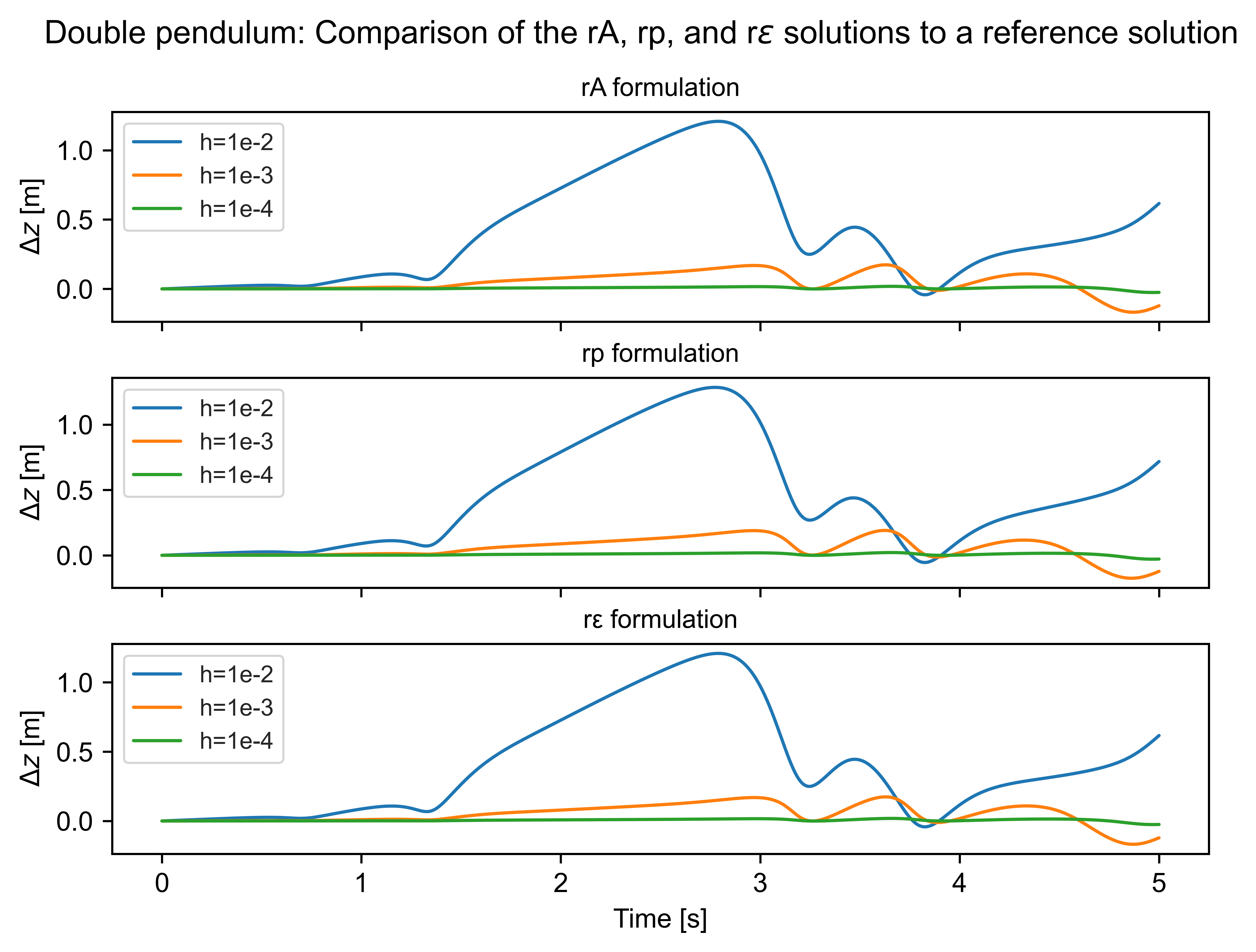}	
		\caption{Double pendulum mechanism, dynamics, two degrees of freedom: Absolute difference (\( \Delta z \)) between the {\rA}, {\rp}, and {\reps} solutions for the position of the z-coordinate of the second pendulum and a reference solution using an ODE solver with a step size of 1e-7. Note that all three formulations trend towards the reference solution as the step size \( h \) decreases with nearly identical \( \Delta z \) values. }\label{fig:solutionDeltas}
	\end{figure}

	To the best of our knowledge, the methodology proposed herein for the {\rA} formulation has not been used elsewhere. To validate the results produced by {\rA}, the dynamics analysis is carried out on the double pendulum for 5 seconds with various step sizes \( h \) and compared to a reference solution that uses an ordinary differential equation solver with a step size \( h \) of \num{1e-7}. As seen in Fig.~\ref{fig:solutionDeltas}, the {\rA} formulation tends towards the reference solution as the step size decreases, and it does so in the same way as {\rp} and {\reps}. 

	\medskip

	\noindent{\textbf{Run Times}}. As shown in Table~\ref{tab:dyn-speed-up}, {\rA} is approximately 2.5 times faster than {\rp} and two times faster than {\reps}. To the best of our knowledge, no similar study is reported in the literature that compares the {\rp} and {\reps} formulations. Surprisingly, {\reps} consistently turned out to be faster than {\rp}. This was unexpected, since the {\rp} formulation was adopted in mid to late 1980s \cite{Nikravesh,Haug89} as an improvement to the {\reps} formulation in use at the time.
	
	\begin{table}[]
		\centering
		\begin{tabular}{rllll}
			\toprule
			& \multicolumn{2}{c}{C1} & \multicolumn{2}{c}{C2} \\
			\addlinespace[-1ex]
			& {\rA}    & \reps{} & {\rA} & \reps{} \\ \midrule
			1-Pendulum   & 2.43 & 1.29 & 2.46  & 1.69 \\
			2-Pendulum   & 2.63 & 1.21 & 2.39  & 1.53 \\
			Four Link    & 2.62 & 1.21 & 2.41  & 1.44 \\
			Slider Crank & 2.93 & 1.31 & 2.29  & 1.30 \\ \bottomrule
		\end{tabular}
		\caption{Speedup, relative to {\rp} formulation, dynamics. Step size of \num{1e-3}, stopping tolerance $ {{\Theta}}=\num{1e-3} $. All times were computed as the average across 10 runs of the simulation, and there was no significant variation in the run-times. Table includes the double pendulum mechanism used to demonstrate the {\rA} solution.}
		\label{tab:dyn-speed-up}
	\end{table}
	
	\medskip
	
\begin{figure}
		\centering
		\includegraphics*[width=\columnwidth]{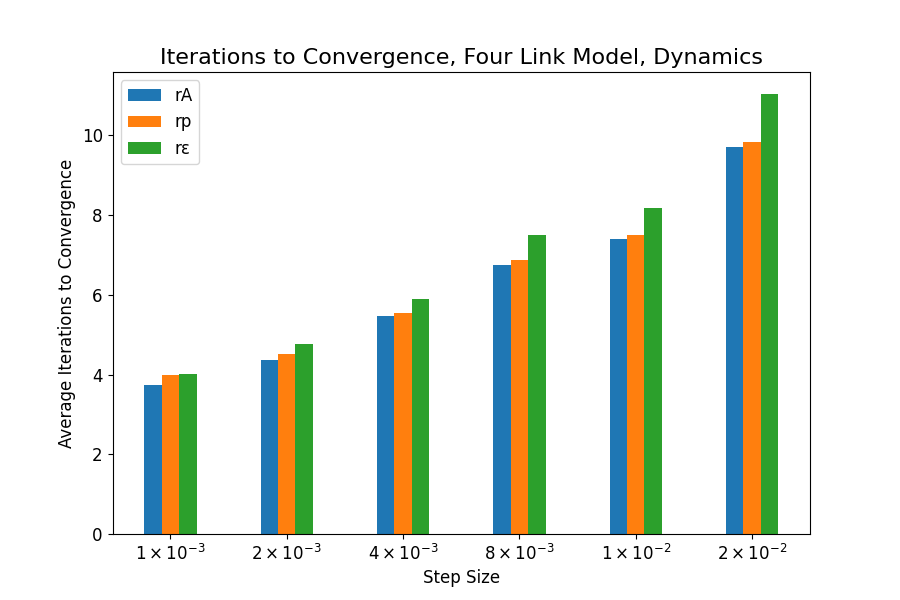}
		\caption{four link mechanism, dynamics: Iterations to convergences. Stopping criteria chosen such that $ {{\Theta}} \; h^2 = 10^{-11} $. Iterations reported are the average across a \SI{3}{s} long analysis. The dependence on step-size is not surprising; note, however, that there is no significant difference between the three formulations.}\label{fig:four link-iter-dyn}
		\vspace{-10pt}
	\end{figure}

	\noindent{\textbf{Iterations to Convergence}}. In the direct index 3 DAE solution approach embraced, at each time step $ t_{n} $, the quantities solved for include: $ {\ddot {\bf r}}_{i,n} $, $ {\dot{\bar {\bm \omega}}}_{i,n} $, and $ {\bm \lambda}_{n} $ in the {\rA} formulation; $ {\ddot {\bf r}}_{i,n} $, $ {{\ddot{\bf p}}}_{i,n} $, and $ {\bm \lambda}_{n} $ in the {\rp} formulation; and $ {\ddot {\bf r}}_{i,n} $, $ {\ddot{{\bm \epsilon}}}_{i,n} $, and $ {\bm \lambda}_{n} $ in the {\reps} formulation. At Newton iteration $ (k) $, a vector containing the body accelerations and the Lagrange multipliers is corrected by a value $ {{\delta}}^{(k)} $, see Eq.~(\ref{eq:corrections}). The iterative process concludes when either the norm of the correction $ {\vect{\delta}}^{(k)} $ is smaller than a threshold value $ {{\Theta}} $, in which case a solution at $ t_n $ was found; or when the iteration count reached a limit number $ K $, in which case the simulation failed:
	\begin{equation}
		\label{eq:stoppingCriteria}
		\| {\vect{\delta}}^{(k)} \|_2 < {{\Theta}} \qquad \mbox{or} \qquad k=K \; .
	\end{equation}
	Note that if the value of the acceleration \textit{correction} is less than $ {{\Theta}}$, then the positions are going to be accurate within $ h^2 {{\Theta}} $, while the velocities are accurate within $ h {{\Theta}} $. Thus, as the step size $ h $ decreases, $  {{\Theta}}$ should be relaxed, since it is not reasonable to expect positions more accurate than, for instance, machine precision. As such, a value of \num{1e-11} is chosen as a value that is reasonable to expect for the accuracy in the positions, and as the step size $ h $ decreases, $ {{\Theta}} $ is chosen such that $ {{\Theta}} \; h^2 = 10^{-11} $. Thus, when $ h=10^{-4} $, $  {{\Theta}} = 10^{-3}$, while when $ h=10^{-3} $, $  {{\Theta}} = 10^{-5}$. The plot in Fig.~\ref{fig:four link-iter-dyn} illustrates that the iterations taken until convergence are strongly dependent on the chosen step-size but do not depend significantly on the formulation used.
	
	\medskip
	
	\noindent{\textbf{Order Analysis}}. The current implementations use a first order implicit Euler integration scheme. The first order accuracy is confirmed with an order analysis conducted as follows: for each of the three solvers ({\rA}, {\rp}, and {\reps}) a dynamics analysis of the model Y$ \in $\{single pendulum, slider crank, four-bar link\} is run for three seconds. A kinematics analysis of Model Y is also run with a tight tolerance to generate ``ground truth'' data (note that all models have zero degrees of freedom owing to prescribed motions). At $ T_{end}=3 $ sec, the state of the system is compared against the ground truth data. The absolute value of the difference between the dynamics results and ground truth is plotted as a function of step size $ h $ on a log-log scale so that the slope of the line illustrates the order of the solver. Note that as the step size decreased, the stopping tolerance was updated according to $ {{\Theta}} \; h^2 = 10^{-11} $ relation (see discussion in \textbf{Iterations to Convergence} above).
	\begin{figure}
		\centering
		\includegraphics*[width=\columnwidth]{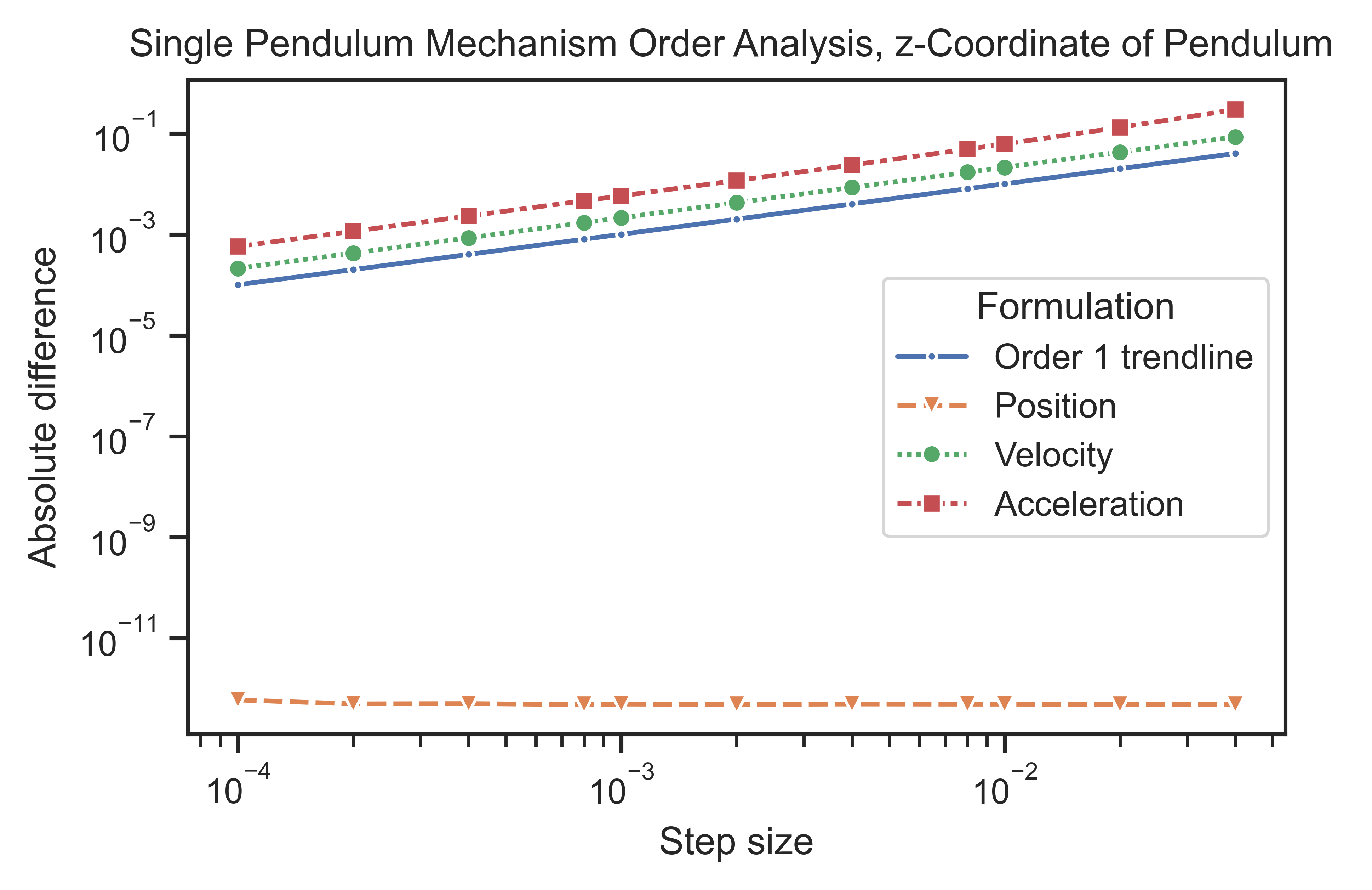}
		\caption{Order analysis of the {\rA} formulation applied to the single pendulum model; backward Euler; $ z $ coordinate of the pendulum.}\label{fig:rA-convSinglePend}
	\end{figure}
	
	The order analysis results for the single pendulum mechanism using {\rA} are shown in Fig.~\ref{fig:rA-convSinglePend}. As expected, the translational velocity and acceleration errors in the $ z $-component after 3 seconds of dynamics are parallel to the blue "order 1 trendline". Note that there is no error in the position-level data since the index 3 solution approach explicitly enforces the position (but not acceleration or velocity) kinematic constraint equations and the simple motion is prescribed on the pendulum. For that reason, this data is left out in the remaining order analysis plots. 		
	
\begin{figure}
		\centering
		\includegraphics*[width=\columnwidth]{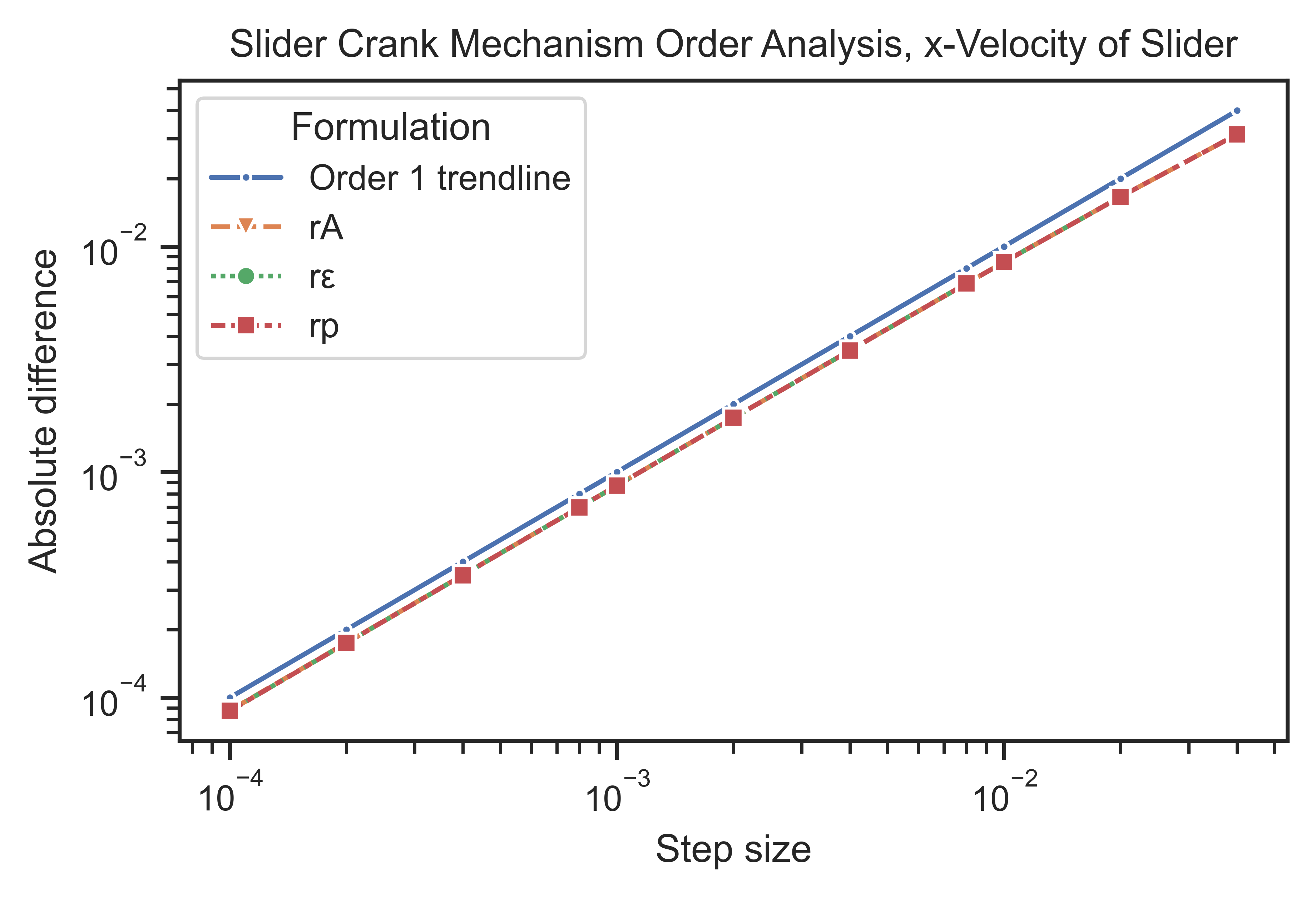}
		\includegraphics*[width=\columnwidth]{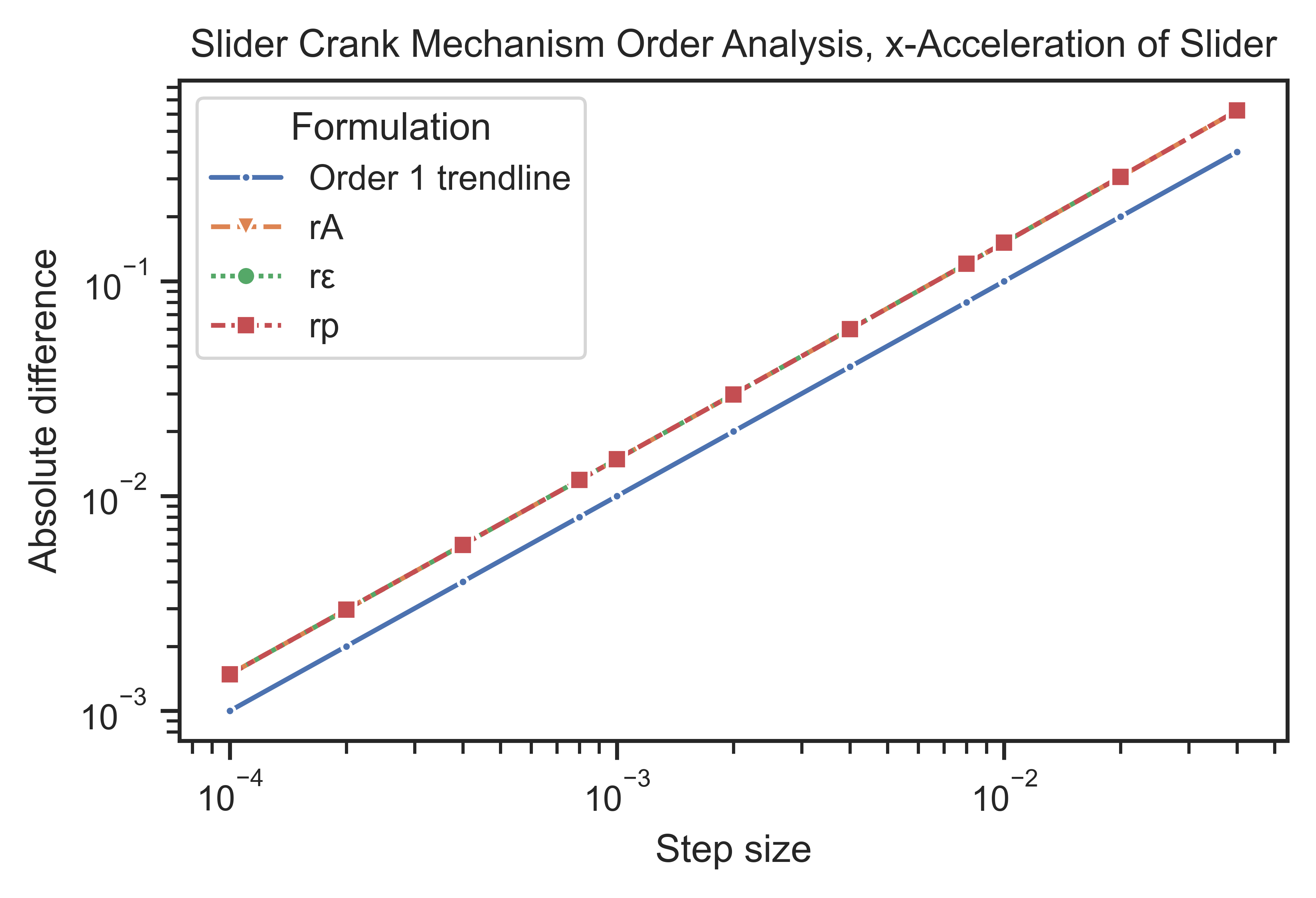}
		\caption{Sample order analysis of the three formulations for the slider crank mechanism; backward Euler; $ x $ coordinate of the slider.}\label{fig:SliderCrankOrderAnalysis-velocity}
	\end{figure}

\begin{figure}
		\centering
		\includegraphics*[width=\columnwidth]{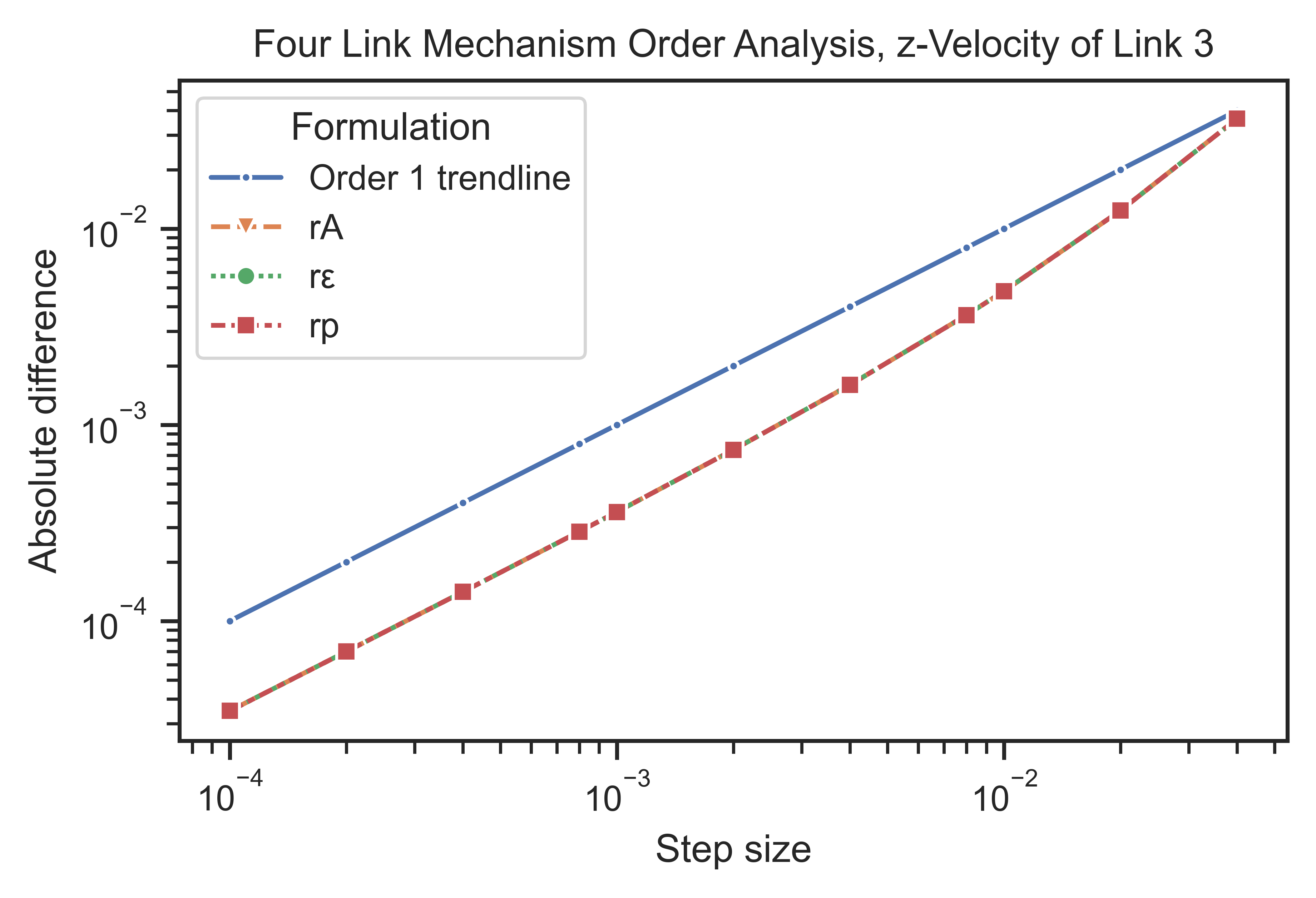}	
		\includegraphics*[width=\columnwidth]{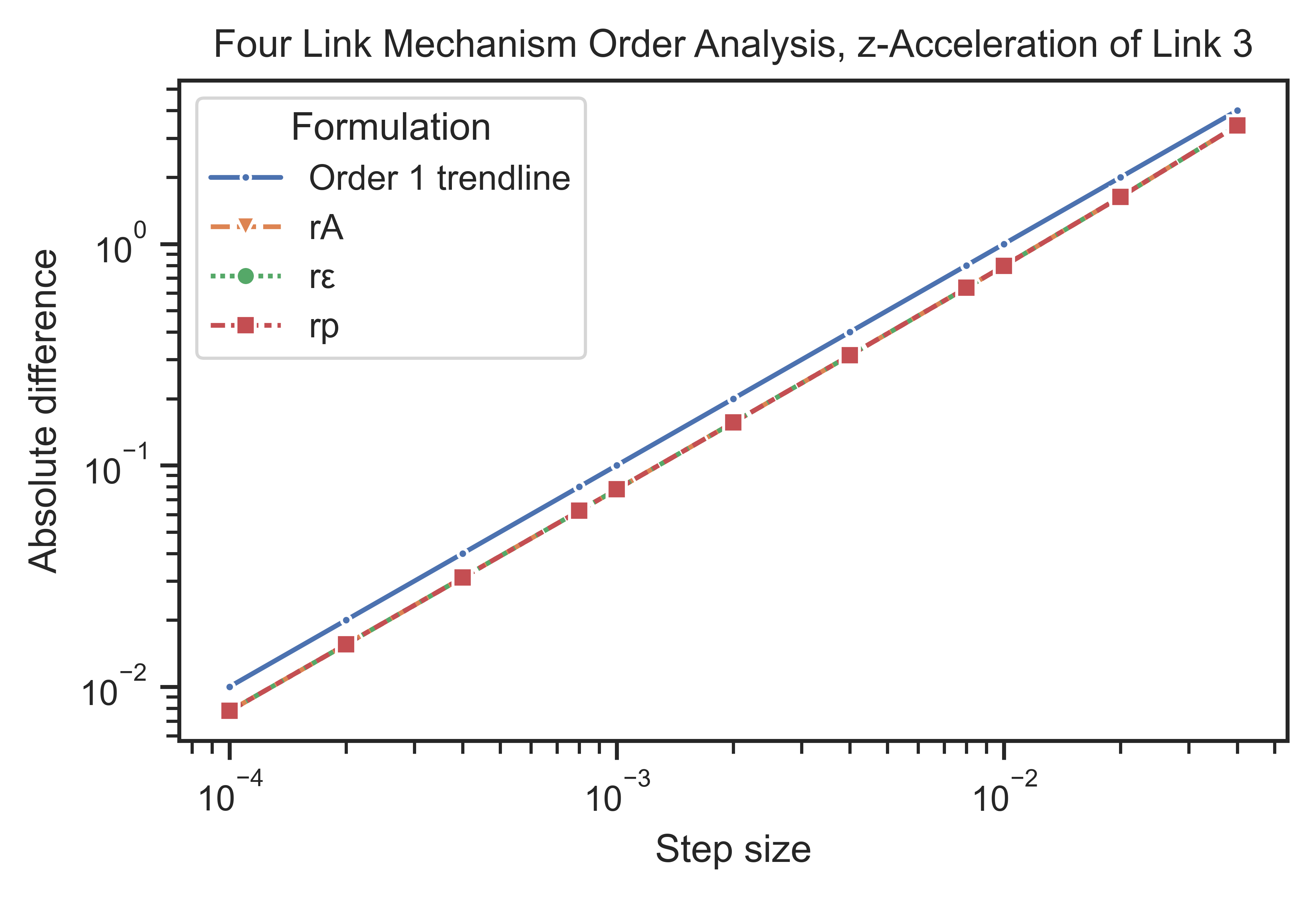} 
		\caption{Sample order analysis of the three formulations for the four link mechanism; backward Euler; $ z $ coordinate of the third link.}\label{fig:FourLinkOrderAnalysis-rArEps}
	\end{figure}

	Figs.~\ref{fig:SliderCrankOrderAnalysis-velocity} and \ref{fig:FourLinkOrderAnalysis-rArEps} represent a sample of data selected from \cite{TR-2020-08-JayAllieDan}, where we report comprehensive order analysis results for {\rA}, {\rp}, and {\reps} in conjunction with the single pendulum, slider crank, and four-bar mechanisms. Therein, results are reported both for C1 and C2; the results reported herein draw exclusively on C1. 
	
	\section{Software, Models, and Data Availability}
	\label{sec:softwareAvailability}
	The models and software used to generate the results reported herein are available in a public repository on GitHub for unfettered inspection, use, and distribution \cite{metadataASME-rA-formulation2021}. The Python software is general purpose; it can be used to simulate arbitrary mechanisms assembled via the four GCONs discussed herein. The model definitions are specified via a json file \cite{json}.
	
	\section{Conclusions and Future Work}
	\label{sec:conclusions}
	Classical multibody dynamics in absolute coordinates commonly relies on Euler parameters or Euler angles to produce the orientation matrix of each body in the system at each time step. The methodology of the {\rA} formulation proposed herein eschews this step by explicitly computing the orientation matrix $ \oriMat $ via numerical integration. The highlight of this contribution is that the methodology discussed herein generalizes this approach to arbitrarily complex multibody systems by systematically formulating lower-pair joints in terms of four basic kinematic geometric constraints: {\textbf{DP1}}, {\textbf{DP2}}, {\textbf{D}}, and {\textbf{CD}}. Understanding the first order variation of these GCONs relative to changes in the orientation of a body draws on the ability to produce the first order variation of two simple quantities: $ \oriMat \sbar$ and $ {\oriMat}^T \vect{s}$, which are used time and again in the kinematic constraint equations. The solution of the Newton-Euler equations is found via a first order implicit Euler integration scheme that numerically integrates the index 3 DAEs of multibody dynamics. The SO(3) structure of the rotation matrices is accounted for by applying an exponential map numerical integration method that builds off Euler's theorem and the Rodrigues formula. 
	
	A performance comparison is made between the proposed {\rA} formulation and two commonly used formulations {\rp} and {\reps} that use Euler parameters and Euler angles as generalized coordinates, respectively. Using {\rp} as a baseline, it is found that {\rA} is approximately two times faster than {\reps}, which, surprisingly, turned out to be faster than {\rp}. 
	
	The behavior of the {\rA} solution approach and the results of the performance comparisons are confirmed by two independently developed publicly available Python codes. The speed gains associated with {\rA} are traced back to a simpler form of the equations of motion and a terser Jacobian matrix. The speed gains associated with {\reps} indicate that the larger linear system in {\rp} is more of a computational bottleneck than the expensive trigonometric functions involved in {\reps}.
	
	In future work, we will investigate how the reported speed gains of {\rA} over the alternative {\rp} or {\reps} solutions can be leveraged in other contexts. First, we plan to investigate its performance in conjunction with a second order Lie group implicit integrator. It also remains to formulate an {\rA} solution approach using a state space method, which reduces the DAE to an ODE problem \cite{Wehage82,betschNullSpace2006}. Another direction of investigation is tied to handling friction and contact in a differential-variational framework \cite{negrutSerbanTasoraJCND2017}. The handling of higher-pair kinematic constraints will require additional insights since the four GCONs introduced will not be sufficient to capture, for instance, unilateral kinematic constraints. Finally, it remains to investigate how the simpler form of the equations of motion impacts the controls problem in multibody dynamics as well as the task of Machine Learning, where fast simulation is key for effective training.

	
	\bibliographystyle{unsrt}
	\bibliography{BibFiles/refsMBS,BibFiles/refsChronoSpecific,BibFiles/refsDEM,BibFiles/refsFSI,BibFiles/refsTerramech,BibFiles/refsSBELspecific,BibFiles/refsRobotics,BibFiles/refsCompSci,BibFiles/refsLinAlgebra,BibFiles/refsNumericalIntegr}

\end{document}